\documentclass[aps,showpacs,amsmath,amssymb,twocolumn,nofootinbib]{revtex4-1}

\usepackage{cancel}
\usepackage{cases}
\usepackage{subfigure}
\usepackage{graphicx}
\usepackage{dcolumn}
\usepackage{bm}
\usepackage{color}
\usepackage[colorlinks=true,pdfstartview=FitV,linkcolor=blue,citecolor=blue,urlcolor=blue]{hyperref}
\usepackage[mathlines]{lineno}
\usepackage[dvipsnames]{xcolor}
\usepackage{amsmath}
\usepackage{ytableau}
\usepackage{adjustbox}

\usepackage{textcase}
\usepackage{colortbl}
\usepackage{cancel}
\usepackage{ulem}
\usepackage{soul}

\everymath{\displaystyle}
\begin{document}
\begin{flushright}
CERN-TH-2022-077
\end{flushright}
\newcommand{\beq}{\begin{equation}}
\newcommand{\eeq}{\end{equation}}
\newcommand{\beqs}{\begin{eqnarray}}
\newcommand{\eeqs}{\end{eqnarray}}

\title{PBH assisted search for QCD axion dark matter}

\author{Gongjun Choi$^{1}$}
\thanks{{\color{blue}gongjun.choi@cern.ch}}

\author{and Enrico D. Schiappacasse$^{2,3,4}$}
\thanks{{\color{blue} enrico.schiappacasse@helsinki.fi}}

\affiliation{$^{1}$ Theoretical Physics Department, CERN, CH-1211 Gen\`eve 23, Switzerland}

\affiliation{$^{2}$ Department of Physics, University of Jyv$\ddot{a}$skyl$\ddot{a}$, P.O.Box 35 (YFL), FIN-40014 Jyv$\ddot{a}$skyl$\ddot{a}$,  Finland}

\affiliation{$^{3}$ Helsinki Institute of Physics, University of Helsinki, P.O. Box 64, FIN-00014 Helsinki, Finland}

\affiliation{$^{4}$ Department of Physics and Astronomy, Rice University, Houston, TX, 77005, U.S.A.}

\date{\today}

\begin{abstract}
The entropy production prior to BBN era is one of ways to prevent QCD axion with the decay constant $F_{a}\in[10^{12}{\rm GeV},10^{16}{\rm GeV}]$ from overclosing the universe when the misalignment angle is $\theta_{\rm i}=\mathcal{O}(1)$. As such, it is necessarily accompanied by an early matter-dominated era (EMD) provided the entropy production is achieved via the decay of a heavy particle. In this work, we consider the possibility of formation of primordial black holes during the EMD era with the assumption of the enhanced primordial scalar perturbation on small scales ($k>10^{4}{\rm Mpc}^{-1}$). In such a scenario, it is expected that PBHs with axion halo accretion develop to ultracompact minihalos (UCMHs). We study how UCMHs so obtained could be of great use in the experimental search for QCD axion dark matter with $F_{a}\in[10^{12}{\rm GeV},10^{16}{\rm GeV}]$. 

\end{abstract}

\maketitle
\section{Introduction}  
The charm of QCD axion lies in the fact that it could not only explain the null observation of CP violation in the strong sector in the Standard Model (SM), but also serve as a good dark matter (DM) candidate. As a pseudo Nambu-Goldstone boson resulting from the spontaneous breaking of the global $U(1)$ anomalous with respect to $SU(3)_{c}$ in the SM, its potential is generated by the instanton-induced fermionic determinantal interaction when QCD becomes non-perturbative. Thereby the axion mass ($m_{a}$) and its decay constant ($F_{a}$) is subject to the relation $m_{a}F_{a}\sim m_{\pi}F_{\pi}\simeq(100{\rm MeV})^{2}$ with $m_{\pi}$ and $F_{\pi}$ the pion mass and decay constant.

Concerning the decay constant, the lower bound based on the stellar cooling process reads $\sim10^{9}{\rm GeV}$~\cite{Raffelt:2006cw}. For a QCD axion model where the axion is the DM candidate and the initial misalignment angle is $\theta_{\rm i}\equiv a_{i}/F_{a}=\mathcal{O}(1)$, how high $F_{a}$ could be is determined by the relic abundance of the axion today ($\Omega_{a}h^{2}$). And $\Omega_{a}h^{2}=\Omega_{\rm DM}h^{2}$ holds for $F_{a}\simeq10^{12}{\rm GeV}$ when the standard cosmological history without the early matter-dominated (EMD) era is assumed. QCD axion window so obtained $10^{9}{\rm GeV}\lesssim F_{a}\lesssim10^{12}{\rm GeV}$ indicates the mass range $\mathcal{O}(10^{-6}){\rm eV}\leq m_{a}\leq\mathcal{O}(10^{-3}){\rm eV}$ for the QCD axion.

The upper bound of $F_{a}$ above, however, can be relaxed provided there is a mechanism either to deplete the relic abundance of axion or to induce $\theta_{\rm i}<\!\!<\mathcal{O}(1)$ before oscillation starts. For the former, as an example, when the axion is coupled to the hidden photon, a rapid axion energy conversion thanks to tachyonic instability in dark photon production could allow for the higher $F_{a}$ than $10^{12}{\rm GeV}$~\cite{Agrawal:2017eqm,Kitajima:2017peg}. For the later, one may consider the possibility of an enhanced axion mass during inflation which results in a suppressed $\theta_{\rm i}$ after inflation~\cite{Dvali:1995ce,Co:2018phi,Buen-Abad:2019uoc}.

The assumption of the entropy production caused by the late time decay of a heavy particle is another way of the energy depletion.\footnote{PBH evaporation before BBN can be also a source of the entropy production~\cite{Bernal:2021bbv,Bernal:2022pue}.} This is particularly convincing possibility when an axion model is considered with supersymmetry (SUSY) or in light of string theory~\cite{deCarlos:1993wie,Banks:1993en,Conlon:2006tq,Svrcek:2006yi,Choi:2006za,Acharya:2010af}. A heavy saxion and moduli fields are generic prediction thereof and hence the axion cosmology becomes naturally incorporated with sources of the entropy production. Insofar as the decay takes place before $T\simeq5{\rm MeV}$ (the temperature of the thermal bath) is reached, the extra radiation arising from the decay of the heavy particle could dilute the energy density of the universe contributed by axion without spoiling the relic abundance of primordial light elements~\cite{Steinhardt:1983ia,Kawasaki:1995vt,Kawasaki:2015pva}. 

The presence of such a heavy degree of freedom ($\sigma$) often implies the non-standard cosmology featured by an EMD era prior to BBN. Soon after starting oscillation, $\sigma$ behaves as a matter ($\rho_{\sigma}\propto R^{-3}$) so as to start to dominate the energy budget of the universe. Therefore, a hunting strategy for the axion with a large decay constant $F_{a}\in[10^{12}{\rm GeV},10^{16}{\rm GeV}]$ could concern phenomenologies attributable to EMD era. One of these could be the fact that the growth rate of the perturbation is linear, differing from that in radiation dominated era (logarithmic). If primordial fluctuations re-entering the horizon during the EMD era are large enough, relatively more efficient formation of the primordial black holes (PBH) than that in the standard cosmology can be expected.

The constraint on the power spectrum of the primordial curvature perturbation on small scales $k\gtrsim\mathcal{O}(10){\rm Mpc}^{-1}$ is still not as strong as the CMB constraint on larges scales $k\lesssim0.2{\rm Mpc}^{-1}$. Thus, having much larger primordial power spectrum on small scales $k\gtrsim\mathcal{O}(10){\rm Mpc}^{-1}$ than in the CMB scale still remains viable, which could possibly develop to formation of heavy enough PBHs with $M_{\rm PBH}\gtrsim\mathcal{O}(10^{-16})M_{\odot}$ that does not evaporate to date. If so, accretion of the axion halo around PBHs is expected, which produces the so-called ultracompact minihalo (UCMH) heavier than $M_{\rm PBH}$. 

In this paper, motivated by the interest in UV physics-supported axion with the large decay constant $F_{a}=\mathcal{O}(10^{16}){\rm GeV}$ and the EMD era prior to BBN era, we study axion cosmology with UCMH made up of PBH and axion halo. We consider the QCD axion decay constant range $F_{a}\in[10^{12}{\rm GeV},10^{16}{\rm GeV}]$. We shall assume two reheating periods: the reheating due to the inflaton right after the inflation ({\bf rh1}) and the reheating due to the heavy scalar particle $\sigma$ (saxion or moduli) decay prior to BBN era ({\bf rh2}). Due to our interest in the specific experimental ways of probing the scenario, we will discuss two cases in this work: {\bf Case I} with $M_{\rm PBH}\lesssim\mathcal{O}(10^{-6})M_{\odot}$ and {\bf Case II} with $M_{\rm PBH}=\mathcal{O}(1)M_{\odot}$.

In Sec.~\ref{sec:axionabundance}, we review the axion relic abundance in the standard cosmology and shows how it could be modified in the presence of entropy production. In Sec.~\ref{sec:ADM}, we go through the details of the numerical computation for the relic abundance of axion DM in the presence of EMD era and the entropy release from the heavy particle decay. In Sec.~\ref{sec:PBH}, with specification of our assumption for the power spectrum of primordial curvature perturbation, we compute the mass and the fraction of PBH in DM as functions of the heavy particle mass and reheating temperature (rh2). And then in Sec.~\ref{sec:UCMH2}, we explain how the PBH could develop to UCMH. In Sec.~\ref{sec:exp}, we discuss how the large decay constant axion picture can be experimentally probed based on the enhanced axion dark matter local density due to the tidal stream arising from disruption of UCMH characterized by $M_{\rm PBH}\lesssim\mathcal{O}(10^{-6})M_{\odot}$ (case I). In Sec.~\ref{sec:caseII}, we study another way of indirect detection of axions with $F_{a}\in[10^{12}{\rm GeV},5\times10^{13}{\rm GeV}]$ through detection of the transient radio signal generated from encounter between a neutron star and UCMH characterized by $M_{\rm PBH}=\mathcal{O}(1)M_{\odot}$. Finally, in Sec.~\ref{sec:conclusion}, we conclude by summarizing the main points of the scenario and discussing its future outlook. 

\section{Diluting Axion Relic Abundance by Entropy Production}
\label{sec:axionabundance}
Having $F_{a}\in[10^{12}{\rm GeV},10^{16}{\rm GeV}]$ for the QCD axion in mind, PQ-breaking is expected to take place before or during the inflation in the scenario we consider, which precludes generation of the associated topological defects and axion production from their decay.\footnote{With $A_{s}=2.1\times10^{-9}$, ($68\%$ C.L., Planck TT,TE,EE+lowE+lensing)~\cite{Planck:2018jri} and $r<0.036$ ($95\%$ C.L., BICEP/Keck)~\cite{BICEP:2021xfz}, the Gibbons-Hawking temperature $T_{I}=H_{I}/2\pi=M_{P}\sqrt{A_{s}r/8}$ is constrained to be $T_{I}\lesssim8\times10^{12}{\rm GeV}$ where $M_{P}=2.4\times10^{18}{\rm GeV}$ is the reduced Planck mass.} This makes the misalignment mechanism the dominant non-thermal axion production channel aside from the negligible thermal production from the thermal plasma via particle scatterings and decays~\cite{Preskill:1982cy,Dine:1982ah,Abbott:1982af}.

In the standard cosmology without an EMD era, when the Hubble expansion rate becomes comparable to the axion mass ($m_{a}$), the axion field starts coherent oscillation. Given that the axion energy density ($\rho_{a}$) scales as $\rho_{a}\propto R^{-3}$ with $R$ the scale factor and the number density ($n_{a}$) is $\rho_{a}=m_{a}n_{a}$, the comoving number density 
($Y_{a}\equiv n_{a}/s$) of axion becomes conserved since then because the entropy density $s$ scales as $s\propto R^{-3}$. 

Depending on whether the oscillation gets started before and after the QCD phase transition, the axion relic abundance expression could be different due to the different oscillation temperature $T_{\rm osc}\simeq\sqrt{m_{a}(T_{\rm osc})M_{P}}$~\cite{Fox:2004kb}. For $F_{a}\lesssim2\times10^{15}{\rm GeV}$ ($T_{\rm osc}>\Lambda_{\rm QCD}$), the axion relic abundance from the misalignment mechanism reads
\beq
\Omega_{a}h^{2}\simeq(2\times10^{4})\times\theta_{\rm i}^{2}\times\left(\frac{F_{a}}{10^{16}{\rm GeV}}\right)^{\frac{7}{6}}\,,
\label{eq:Omegaa1}
\eeq
where $\theta_{\rm i}\equiv a_{\rm i}/F_{a}$ is the initial misalignment angle. In contrast, for $F_{a}\gtrsim2\times10^{17}{\rm GeV}$ ($T_{\rm osc}<\Lambda_{\rm QCD}$), it reads 
\beq
\Omega_{a}h^{2}\simeq(5\times10^{3})\times\theta_{\rm i}^{2}\times\left(\frac{F_{a}}{10^{16}{\rm GeV}}\right)^{\frac{3}{2}}\,.
\label{eq:Omegaa2}
\eeq
Here $\Omega_{a}h^{2}$ for the decay constant lying in an intermediate region $F_{a}=\mathcal{O}(10^{16}){\rm GeV}$ concerns strong QCD effects, which makes both Eq.~(\ref{eq:Omegaa1}) and (\ref{eq:Omegaa2}) fail to apply. At the moment, for $F_{a}\in[10^{12}{\rm GeV},10^{16}{\rm GeV}]$ which is our main interest, let us refer to Eq.~(\ref{eq:Omegaa1}).

As we envision in our scenario, when the current DM is identified with the axion, $\Omega_{a}h^{2}=\Omega_{\rm DM}h^{2}\simeq0.12$ demands
\beq
\theta_{\rm i}\lesssim2.45\times10^{-3}\times\left(\frac{F_{a}}{10^{16}{\rm GeV}}\right)^{-\frac{7}{12}}\,,
\label{eq:thetai}
\eeq
where the bound is saturated when the axion explains the whole population of DM. From Eq.~(\ref{eq:thetai}), one can infer that $\theta_{\rm i}\lesssim0.5$ is required for $F_{a}\gtrsim10^{12}{\rm GeV}$ and particularly the string axion with $F_{a}=\mathcal{O}(10^{16}){\rm GeV}$ can easily exceed the current DM relic density unless a mechanism is assumed to naturally account for the sufficiently small initial misalignment angle $\theta_{\rm i}\leq\mathcal{O}(10^{-3})$.

On the other hand, of course, provided axions somehow experience depletion in energy after they start the oscillation (and decouple from the thermal plasma for the thermal axion component), $\Omega_{a}h^{2}>\Omega_{\rm DM}h^{2}$ can be avoided even with $\theta_{\rm i}=\mathcal{O}(1)$ and $F_{a}>10^{12}{\rm GeV}$. As a matter of fact, this is not really a contrived set-up if the low energy story of axion is embedded in either supersymmetric models or string theory. These theories naturally accommodating heavy moduli fields, the presence of the EMD era and the following entropy production through the decay of the heavy particle could enable depletion in axion abundance.

Along this line of reasoning, we consider the large decay constant axion DM scenario in which $\Omega_{a}h^{2}\lesssim\Omega_{\rm DM}h^{2}$ is achieved even with $\theta_{\rm i}=\mathcal{O}(1)$ thanks to the dilution by the entropy production from the moduli field ($\sigma$) decay. Below we discuss $\Omega_{a}h^{2}$ schematically for illustration which will be improved later in Sec.~\ref{sec:ADM} based on the detailed numerical computation. As a concrete example of $\sigma$, in this paper, we consider a heavy moduli field with the decay rate
\beqs
\Gamma_{\sigma}&=&c\frac{m_{\sigma}^{3}}{M_{P}^{2}}\cr\cr&\simeq&c\times1.69\times10^{-22}{\rm GeV}\times\left(\frac{m_{\sigma}}{100{\rm TeV}}\right)^{3}\,,
\label{eq:saxiondecayrate}
\eeqs
where $c$ is a model dependent parameter (see, e.g. ~\cite{Cicoli:2016olq}). We choose $c=1$ for the analysis from here on. For the case with other choice of $c$, our result can be applied with the proper re-scaling of $m_{\sigma}$. 

Let us define $s_{\rm old}$ to be the (existing) entropy density before the axion oscillation starts and $s_{\rm new}$ to be the total entropy density including both the existing entropy density and new one released from the decay of $\sigma$. Then we can quantify the amount of the entropy production by the ratio
\beq
\Delta\equiv\frac{S_{\rm new}}{S_{\rm old}}=\frac{s_{\rm new}(R_{\rm rh2})R_{\rm rh2}^{3}}{s_{\rm old}(R_{\rm osc})R_{\rm osc}^{3}     }=\frac{s_{\rm new}(R_{\rm rh2})}{s_{\rm old}(R_{\rm rh2})}\,,
\label{eq:Delta}
\eeq
where the upper case $S=sR^{3}$ denotes the entropy. Now in terms of $\Delta$ in Eq.~(\ref{eq:Delta}), we can rewrite the axion relic abundance today as
\beqs
\Omega_{a}h^{2}&=&\frac{m_{a}s_{0}h^{2}}{\rho_{\rm cr,0}}\left(\frac{n_{a}(R_{\rm rh2})}{s_{\rm new}(R_{\rm rh2})}\right)\cr\cr
&=&\frac{m_{a}s_{0}h^{2}}{\rho_{\rm cr,0}}\left(\frac{n_{a}(R_{\rm rh2})}{s_{\rm old}(R_{\rm rh2})}\right)\left(\frac{s_{\rm old}(R_{\rm rh2})}{s_{\rm new}(R_{\rm rh2})}\right)\cr\cr
&=&\frac{m_{a}s_{0}h^{2}}{\rho_{\rm cr,0}}\left(\frac{n_{a}(R_{\rm osc})}{s_{\rm old}(R_{\rm osc})}\right)\Delta^{-1}\cr\cr
&=&\Omega_{a, {\rm old}}h^{2}\Delta^{-1}
\label{eq:Omegaanew}
\eeqs
where we defined $\Omega_{a, {\rm old}}h^{2}$ to be the would-be axion relic abundance in the absence of the entropy production which is therefore identified with Eq.~(\ref{eq:Omegaa1}). Here $s_{0}$ and $\rho_{\rm cr,0}$ are the current entropy density and the critical energy density respectively, and $h$ parametrizes the Hubble expansion rate today through $H_{0}=100h{\rm km/sec/Mpc}$.

From Eq.~(\ref{eq:Omegaanew}), for example, regarding the string axion with $F_{a}=\mathcal{O}(10^{16}){\rm GeV}$, it can be realized that $\Omega_{a}h^{2}\lesssim\Omega_{\rm DM}h^{2}$ holds for $\Delta=\mathcal{O}(10^{5})$ even for $\theta_{\rm i}=\mathcal{O}(1)$. Note that the energy conversion of $\sigma$ to the new radiation (the entropy production) gives us the relation
\beq
\frac{4\rho_{\sigma}(R_{\rm rh2})}{3T_{\rm rh2}}=s_{\rm new}(R_{\rm rh2})\,,
\label{eq:srho}
\eeq
where $R_{\rm rh2}$ can be read from $H(R_{\rm rh2})\simeq\Gamma_{\sigma}$ with $\Gamma_{\sigma}$ the decay rate of $\sigma$. Therefore, as far as the heavy particle energy is large enough on decay so as to guarantee large enough $\Delta$, the string axion scenario, not to mention $F_{a}\in[10^{12}{\rm GeV},10^{16}{\rm GeV}]$, can be saved from disastrous overclosure of the universe.

We conclude this section by commenting on a upper bound on the inflation scale in the large decay constant axion scenario with the EMD era. During the inflation, the axion is essentially massless free scalar and thus its perturbation is subject to the CMB constraint on the isocurvature mode $P_{\rm iso}<0.038\times(2\times10^{-9})$ at $k_{\rm pivot}=0.05{\rm Mpc}^{-1}$~\cite{Planck:2018jri}, i.e.,
\beq
P_{\rm iso}\simeq\left(\frac{\Omega_{a}h^{2}}{\Omega_{\rm CDM}h^{2}}\right)^{2}\left(\frac{H_{I}}{\pi F_{a}\theta_{\rm i}}\right)^{2}<0.038\times(2\times10^{-9})\,,
\label{eq:Piso}
\eeq
where $H_{I}$ is the Hubble expansion rate during the inflation. This in turn gives the upper bound on $H_{I}$ as follows
\beq
H_{I}\lesssim2.7\times10^{10}{\rm GeV}\left(\frac{\theta_{\rm i}}{0.1}\right)\left(\frac{F_{a}}{10^{16}{\rm GeV}}\right)\left(\frac{\Omega_{\rm CDM}h^{2}}{\Omega_{a}h^{2}}\right)\,.
\label{eq:Hinf}
\eeq
For a given $\theta_{\rm i}$ and $F_{a}\in[10^{12}{\rm GeV},10^{16}{\rm GeV}]$, we shall assume an inflationary dynamics with a low enough $H_{I}$ complying with Eq.~(\ref{eq:Hinf}) from hence.


\section{Axion Dark Matter with Early Matter Dominated era}
\label{sec:ADM}
The evolution equation for the axion field in the early universe is given by
\begin{equation}
\left(\partial_t^2 + 3H(t)\partial_t -R^{-2}(t)\nabla_x^2\right)a(x) + V'(a) = 0\,,  
\label{eq:axioneqm}
\end{equation}
where $H(t)$ is the Hubble expansion rate, $a(x)$ is the axion field, ${\bf{x}}$ is the three-vector denoting the co-moving spatial coordinates, and $V(a)$ is the axion effective potential. The potential comes from non-perturbative QCD effects and it may be written as 
\begin{equation}
V(a)=F_a^2 m_a^2(T)\left[ 1 - \text{cos}(a/F_a)\right]\,,    
\end{equation}
where $m_a(0) = (78\, \text{MeV})^2/F_a$ is the axion mass at zero-temperature and~\cite{PhysRevD.78.083507, Nelson:2018via}
\begin{equation}
m_a(T) = \left\lbrace\begin{array}{ll} 
&m_a(0)\,,\hspace{3.25cm}T/\text{GeV}<0.2\, \\ 
&m_a(0)\left(\frac{0.2\,\text{GeV}}{T}\right)^{6.5}\,,\hspace{0.29cm}0.2\,\leq T/\text{GeV}\leq 1\\
&0.018\,m_a(0)\left(\frac{0.2\,\text{GeV}}{T}\right)^{4}\,.\hspace{0.6cm}T/\text{GeV}>1\,
\end{array}\right.  
\label{eq:maT}
\end{equation}

As mentioned before, we work in the scenario at which the PQ symmetry is broken before or during inflation. Then the energy density of topological defects are diluted away by inflation so that the main contribution to the axion abundance comes from the vacuum misalignment associated with the axion zero-modes. Therefore, we have  $a(x)\rightarrow a(t)$ which reduces Eq.~(\ref{eq:axioneqm}) to
\beqs
&&\ddot{a}(t) + 3H(t)\dot{a}(t) + F_am_a^2(T(t))\text{sin}(a/F_a) = 0\cr\cr
&\Rightarrow&\ddot{a}(t) + 3H(t)\dot{a}(t) + m_a^2(T(t))a(t) = 0\,,
\label{eq:axioneqm2}
\eeqs
where the second line is obtained in the limit $a(t)/F_a \ll 1$ and $m_a(T(t))$ indicates the time dependence of the axion mass via  $T=T(t)$. 

As the initial conditions for the axion field and velocity, we choose $\theta_{\rm i} \sim \mathcal{O}(1)$ and $\dot{\theta}_{\rm i} = 0$, respectively. At any temperature $T$ such that $F_a > T \gg T_{\text{osc}}$, where $T_{\text{osc}}\equiv T(t_{\text{osc}})$ is defined via $m_a(t_{\text{osc}})=3H(t_{\text{osc}})$, the axion field is frozen due to Hubble friction. At $t_{\text{osc}}$, the axion field begins to oscillate and the energy density of the axion zero-modes at that time reads 
\begin{equation}
    \rho_a(t_{\text{osc}}) \approx \frac{1}{2} m_a(t_{\text{osc}})^2\theta_{\rm i}^2 F_a^2\,.\label{eq:Y1}
\end{equation}

To calculate the current axion abundance, we need to trace the evolution of $\rho_{a}$ for $t_{\rm osc}<t<t_{\rm rh2}$ by correctly taking into account the entropy production from the heavy particle decay. To do so, we solve the system of
coupled differential equations below including the Friedman Equation and the time evolution equations for energy densities of the radiation ($\rho_{\text{rad}}$) and the heavy scalar particle ($\rho_{\sigma}$) :
\begin{align}
   & \dot{\rho}_{\sigma}(t) + 3H(t)\rho_{\sigma}(t) = -\Gamma_{\sigma}\rho_{\sigma}\,,\label{eq:rhossigma}\\
    &\dot{\rho}_{\text{rad}}(t) + 4H(t)\rho_{\text{rad}}(t) = \Gamma_{\sigma}\rho_{\sigma}\,,\label{eq:rhorad}\\
    &\dot{\rho}_{\text{initial}}(t) + 4H(t)\rho_{\text{initial}}(t) = 0\,,\label{eq:rhoradinit}\\
        &3H(t)^2M_{P}^{2} = \rho_{\sigma}(t) + \rho_{\text{rad}}(t)\,,\label{eq:rhototal}
\end{align}
where we have neglected  the axion contribution to the total energy density in Eq.~(\ref{eq:rhototal}) and $\Gamma_{\sigma}$ in Eq.~(\ref{eq:saxiondecayrate}) is used for Eq.~(\ref{eq:rhossigma}) and (\ref{eq:rhorad}). Eq.~(\ref{eq:rhoradinit}) is  the evolution equation for the radiation in the absence of the heavy scalar decay ($\rho_{\text{initial}}$). We define $t_{*}$ to be the time at which the EMD era begins and satisfies $\rho_{\sigma}(t_*)=\rho_{\text{rad}}(t_*)=\rho_{\text{initial}}(t_*)$.

The evolution of the heavy scalar field is obtained from Eq.~(\ref{eq:rhossigma}) as
$\rho_{\sigma}(t) = \rho_{\sigma}(t_*)R(t)^{-3}\text{exp}[-\Gamma_{\sigma}t]$. For $t_{*} \lesssim t \ll \Gamma_{\sigma}^{-1}$, we have $\rho_{\sigma}(t) \approx 4M_{P}^{2}/(3t^2)$. The total radiation energy density ($\rho_{\text{rad}}$) can be expressed as the sum of the existing radiation energy density ($\rho_{\text{initial}}$) and the new one from the decay of the heavy scalar ($\rho_{\text{gen}}$), i.e.
\beq
\rho_{\text{rad}}(t) = \rho_{\text{initial}}(t) +\rho_{\text{gen}}(t)\,.
\eeq
The time at which  $\rho_{\text{initial}}(t_{\text{equal}}) = \rho_{\text{gen}}(t_{\text{equal}})$ is satisfied reads $t_{\text{equal}} \,\approx\, t_{*}^{2/5}\Gamma_{\sigma}^{-3/5}$ as shown in~\cite{Georg:2017mqk} (Sec. IIB), which is obtained by replacing $\rho_{\sigma}$ in Eq.~(\ref{eq:rhorad}) with the aforementioned $\rho_{\sigma}(t)$.

With account taken of all the above, we proceed to solve the system of equations in Eqs.~(\ref{eq:rhossigma})-(\ref{eq:rhototal})  by using the following initial conditions:
\begin{align}
&(i)\textcolor{white}{ii}\hspace{0.5cm}  t_{*} = \frac{2}{3\times0.765 m_{\sigma}} \left( \frac{M_p}{\sigma_0} \right)^4\,,\label{icond}\\
&(ii)\textcolor{white}{i}\hspace{0.5cm}  \rho_{\sigma}(t_*) = \rho_{\text{initial}}(t_*)=\frac{4 M_p^2}{3t_*^2}\,,\label{iicond}
\end{align}
\begin{figure}[t!]
\centering
\hspace*{-5mm}
\includegraphics[width=0.4\textwidth]{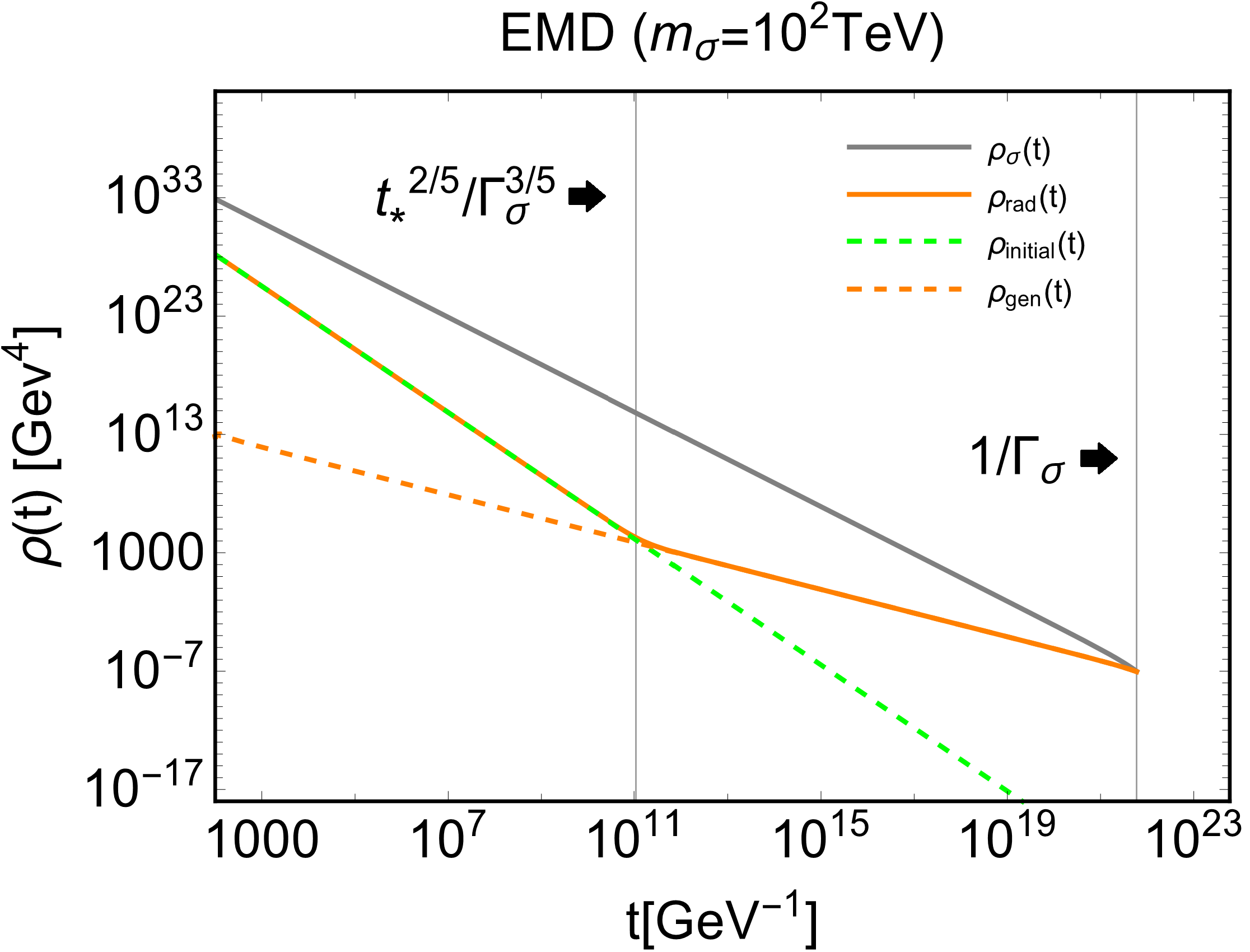}
\includegraphics[width=0.4\textwidth]{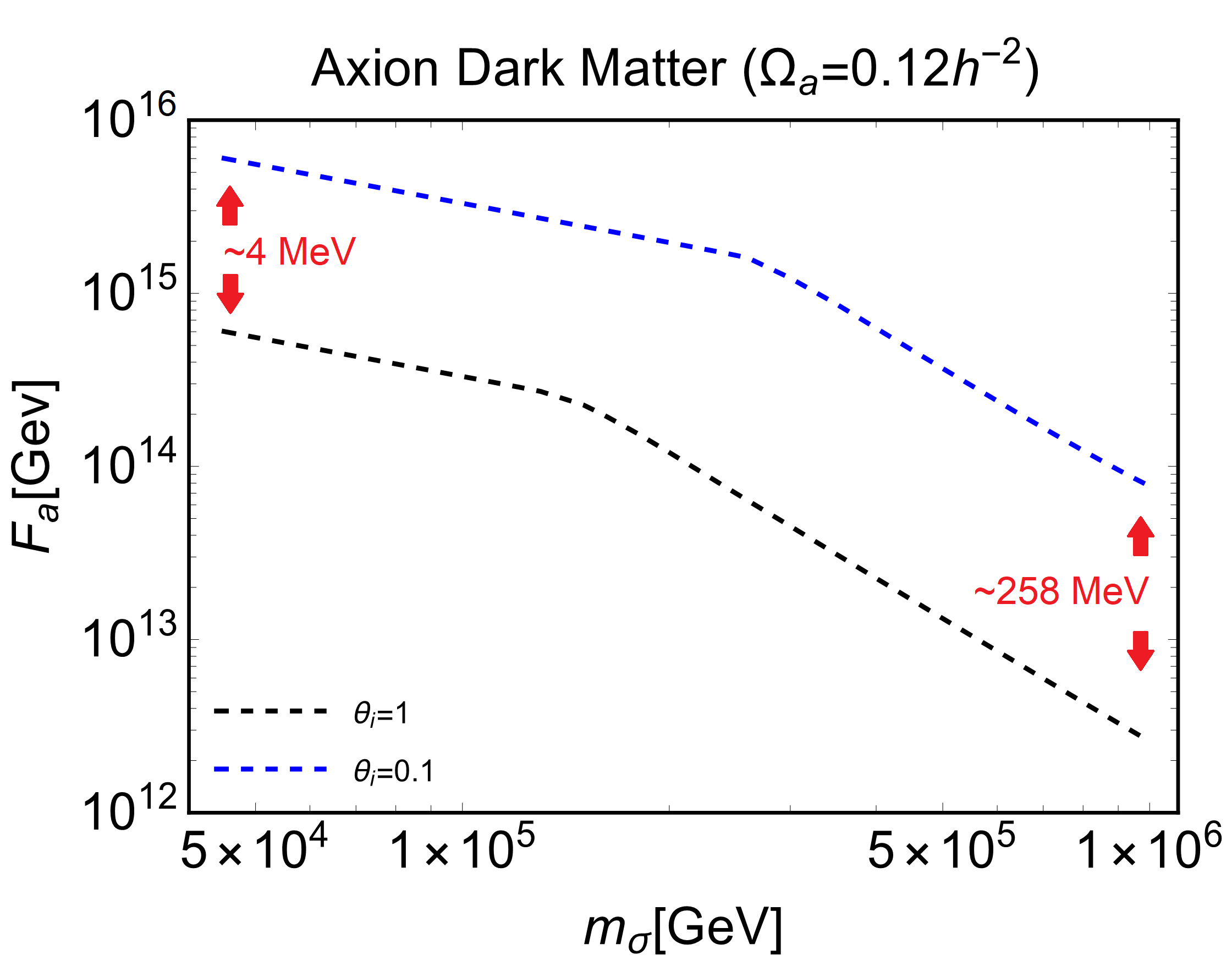}
\caption{(Top) Evolution of energy densities of the heavy scalar $\sigma$ (gray solid) and the radiation (orange solid) during the EMD era with $m_{\sigma}=100\, \text{TeV}$, $\sigma_0=M_p$, and $t_*=8.715\times10^{-6}$ (${\rm GeV}^{-1}$). (Bottom) Values for the axion decay constant $F_{a}$ giving rise to $\Omega_a=0.12 h^{-2}$ for a given heavy scalar mass $m_{\sigma}$ ($\sigma_0=M_p$). Blue (black) dashed line corresponds to the initial misalignment angle $\theta_{\rm i}=0.1$ (1). $T_{\rm rh2}$ associated with $m_{\sigma}=(45,10^3)\,\text{TeV}$ are shown in red color.}
\vspace*{-1.5mm}
\label{fig:EMDevolve}
\end{figure}
\hspace{-0.1cm}where we used Eq.~(\ref{eq:Hinitial}) and $t_* \sim 2/(3H(t_{*}))$.

We map the temperature of the thermal bath $T$ to the cosmological time $t$ via $\rho_{\text{rad}}(t) = (\pi^2/30) g_{*}(T)T^4$ with $g_{*}(T)$ the effective degrees of freedom of the energy density. 

By using Eqs.~(\ref{eq:Y1}) and (\ref{eq:Omegaanew}), we obtain
\begin{equation}
\Omega_a h^2 = \left( \frac{m_a(0)m_a(t_{\text{osc}})\theta_{\rm i}^2F_a^2}{2 \rho_{\text{cr},0}h^{-2}}\right)\left(\frac{s_0 \Delta^{-1}}{s_{\text{old}}(t_{\text{osc}})}\right)\,,
\end{equation}
where $\rho_{\text{cr,0}}=1.053672\times 10^{-5}\, h^2\, \text{GeV/cm}^{3}$
and $s_{\rm old}(t_{\rm osc})$ can be written in terms of $s_{\rm old}(t_{\rm rh2})$ as
 \begin{align}
s_{\text{old}}(t_{\text{osc}}) &=   s_{\text{old}}(t_{\text{rh2}})\left(\frac{R_{\text{rh2}}}{R_{\text{osc}}}\right)^3\,,\\
&=s_{\text{old}}(t_{\text{rh2}}) \left( \frac{\rho_{\sigma}(t_{\text{osc}})}{\rho_{\sigma}(1/\Gamma_{\sigma})} e^{\Gamma_{\sigma}t_{\text{osc}}-1}\right)\,,
\end{align}
where we used the analytic solution of Eq.~(\ref{eq:rhossigma}) and $s_{\text{old}}(t_{\text{rh2}}) \propto \rho_{\text{initial}}^{3/4}(t_{\text{rh2}})$.

In the top panel of Fig.~\ref{fig:EMDevolve}, we show the evolution of different energy densities during the EMD era characterized by $m_{\sigma}=100\,\text{TeV}$. The time estimate at which 
$\rho_{\text{initial}}(t_{\text{equal}}) = \rho_{\text{gen}}(t_{\text{equal}})$ holds, i.e. $t_{\text{equal}} \,\approx\, t_{*}^{2/5}\Gamma_{\sigma}^{-3/5}$ agrees very well with the numerical result. Also, numerically the end of the EMD era can be found in Eq.~(\ref{eq:rhototal}) by requiring $\rho_{\sigma}(\Gamma_{\sigma}^{-1}) \approx \rho_{\text{rad}}(\Gamma_{\sigma}^{-1})$. We confirmed that the numerically obtained $T_{\rm rh2}$ ($t_{\rm rh2}$) is in good agreement with the estimate below from $H(1/\Gamma_{\sigma}) \approx \Gamma_{\sigma}$
\begin{equation}
T_{\text{rh2}} \approx 5\, \text{MeV} \left( \frac{10.75}{g_{*}(5\,\text{MeV})} \right)^{1/4}\left( \frac{m_{\sigma}}{45\,\text{TeV}}\right)^{3/2}\,.
\label{eq:Trh2}
\end{equation}

The bottom panel of Fig.~\ref{fig:EMDevolve} shows values of $F_{a}$ enabling the axion to satisfy $\Omega_{a}h^{2} = 0.12$ today for each given $m_{\sigma}$. In showing the relation obtained based on the numerical computation, we choose two particular values of the initial misalignment angle, $\theta_{\rm i} = 1$ (black dashed line) and $\theta_{\rm i} = 0.1$ (blue dashed line). Note that the entropy injection parametrized by $\Delta$ in Eq.~(\ref{eq:Delta}) allows $F_{a}$ to reach larger values than those in the standard scenario for the shown range of $m_{\sigma}$. The break in both curves separates the region where the axion mass settles down to its zero temperature value from that where the axion mass is temperature-dependent,
as shown in Eq.~(\ref{eq:maT}).  

The reason for seeing such a $F_{a}$'s dependence on $m_{\sigma}$ in Fig.~\ref{fig:EMDevolve} is what follows. Assuming that $\sigma$-field starts oscillation at $t=t_{\rm rh1}$, we see that the ratio $\rho_{\sigma}(R_{\rm rh2})/s_{\rm old}(R_{\rm rh2})\propto T_{\rm rh1}$. Therefore, using the definition of $\Delta$ in Eq.~(\ref{eq:Delta}), we obtain $\Delta\propto(T_{\rm rh1}/T_{\rm rh2})$ because of $\rho_{\sigma}(R_{\rm rh2})/s_{\rm new}(R_{\rm rh2})\propto T_{\rm rh2}$. Eventually combined with Eq.~(\ref{eq:Trh2}), $\Delta\propto T_{\rm rh2}^{-1}$ yields $\Delta\propto m_{\sigma}^{-3/2}$. Now that a larger $F_{a}$ requires a larger dilution factor $\Delta$ to be consistent with $\Omega_{a}h^{2}=0.12$ via Eqs.~(\ref{eq:Omegaa1}) and (\ref{eq:Omegaa2}), it should correspond to a smaller $m_{\sigma}$, which is well reflected in Fig.~\ref{fig:EMDevolve}.


\section{Primordial Black Hole Formation}
\label{sec:PBH}
One of interesting possibilities resulting from the presence of the EMD era is formation of PBHs. For modes of the primordial fluctuation re-entering the horizon during the EMD era, if primordial fluctuations were large enough, these could possibly give rise to efficient formation of PBHs. As a matter of fact, indeed, the existing constraints on the power spectrum of the primordial curvature perturbation, $P_{\zeta}(k)$, for scales $k>10\,{\rm Mpc}^{-1}$ are not as strong as that Planck CMB data provides for $k\lesssim0.2\,{\rm Mpc}^{-1}$.

Interests in PBH as a DM candidate have triggered many studies on ways to enhance the primordial perturbation on small scales. In the simplest minimal scenarios of the single field slow roll inflation, for instance, the presence of a small bump (or dip)~\cite{Hertzberg:2017dkh,Mishra:2019pzq}, and an inflection point in the inflaton potential~\cite{Garcia-Bellido:2017mdw,Ballesteros:2017fsr,Kawasaki:2016pql,Germani:2017bcs,Bhaumik:2019tvl,Ballesteros:2020qam} can induce a peak in the small scale regime of the primordial power spectrum essentially through the ultra-slow roll at a late stage of inflation.\footnote{Large scalar fluctuations on small scales can be also induced in the multifield inflation models~\cite{Braglia:2020eai,Fumagalli:2020adf}, models assuming the coupling between inflaton and gauge fields~\cite{Linde:2012bt,Bugaev:2013fya,Domcke:2017fix}, hybrid inflation models~\cite{Garcia-Bellido:1996mdl,Lyth:2011kj,Bugaev:2011wy,Clesse:2015wea,Kawasaki:2015ppx} and axion monodromy inflation~\cite{Ballesteros:2019hus}.}

Assuming an inflaton potential featured by one of those, one may approximate $P_{\zeta}(k)$ on small scales as a lognormal form to get~\cite{Gow:2020bzo,Bhattacharya:2021wnk}
\beqs
P_{\zeta}(k)&=&A_{s}\left(\frac{k}{k_{\rm CMB}}\right)^{n_{s}-1}\,+\,A_{p}e^{-\frac{(N_{k}-N_{p})^{2}}{2\sigma_{p}^{2}}}\cr\cr
&\simeq&A_{s}\left(\frac{k}{k_{\rm CMB}}\right)^{n_{s}-1}\,+\,A_{p}e^{-\frac{(\log(k/k_{p}))^{2}}{2\sigma_{p}^{2}}}\,,
\label{eq:Pzeta}
\eeqs
where $A_{s}=2.1\times10^{-9}$, $n_{s}=0.965$, $k_{\rm CMB}=0.05\,{\rm Mpc}^{-1}$, and $N_{k}=\log(R(k)/R_{\rm end})$ and $N_{p}=\log(R(k_{p})/R_{\rm end})$ are defined. For the second equality, we used $H_{\rm inf}\simeq{\rm constant}$ during inflation, and $k=R(k)H_{\rm inf}$ and $k_{p}=R(k_{p})H_{\rm inf}$.

Once armed with such $P_{\zeta}(k)$ enhanced at small scales, the moduli-driven EMD era can serve as the environment facilitating formation of PBHs. Unlike in the radiation-dominated (RD) era, density fluctuations grow as $\delta\propto R$ in the matter-dominated era so as to reach the non-linear regime ($\delta=\mathcal{O}(1)$) rather quickly. This is basically due to the absence of the pressure preventing the gravitational collapse of an overdensity to a PBH during the matter-dominated era. With the PBH production probability defined as $\beta\equiv\rho_{\rm PBH}/\rho_{\rm tot}$ evaluated at the time of PBH formation, $\beta_{\rm MD}\simeq0.056\,\sigma^{5}$ was obtained for $\sigma<0.01$ during the matter-dominated era with account taken of suppression in PBH production by the nonspherical effect~\cite{Harada:2016mhb}.\footnote{Note that during a RD era, the production probability is given by $\beta_{\rm RD}(M)\simeq{\rm erfc}[\delta_{c}/\sqrt{2}\sigma(M))]$ with $\delta_{c}\simeq0.414$ and erfc the complementary error function~\cite{Carr:1975qj,Harada:2013epa} (Press-Schechter formalism). It is immediately realized that the production probability is much large for EMD era than RD era.} Here $\sigma$ is the variance of the density fluctuation.

\begin{figure*}[htp]
  \centering
  \hspace*{-5mm}
  \subfigure{\includegraphics[scale=0.4]{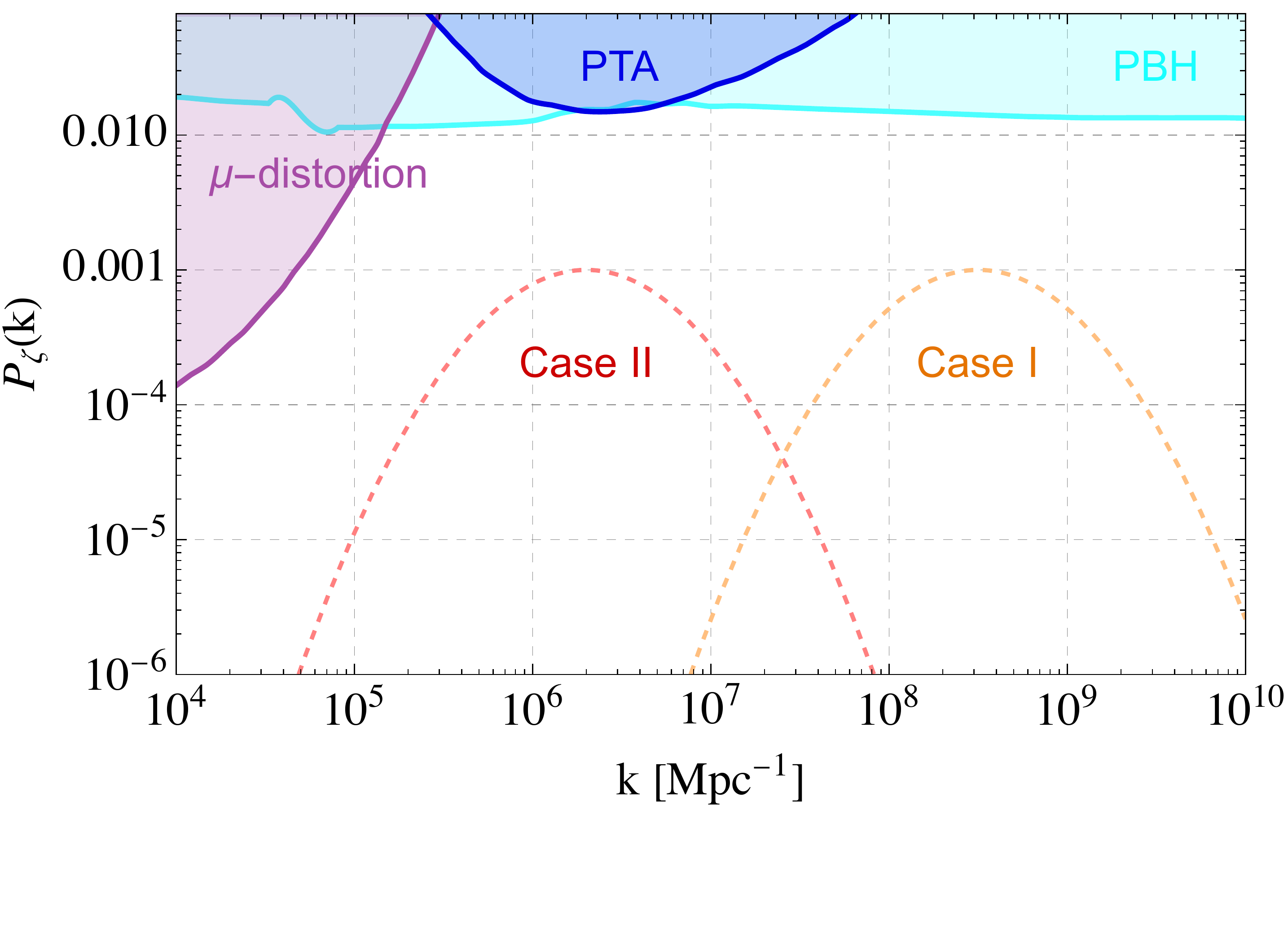}}
  \subfigure{\includegraphics[scale=0.4]{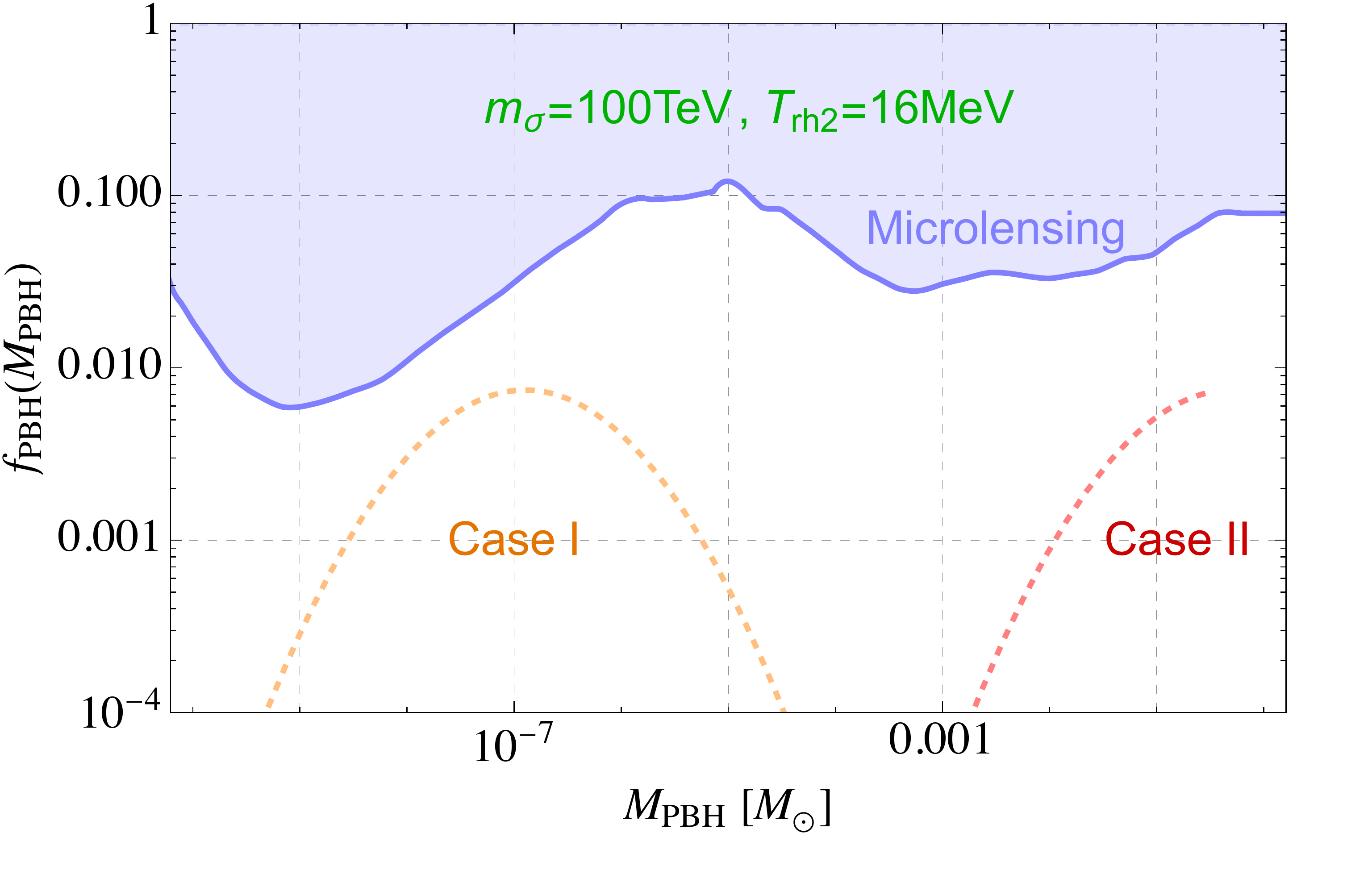}}
  \caption{{\bf Top:} The power spectra of the primordial curvature perturbation on small scales for the case I with $k_{p}=10^{8.5}{\rm Mpc}^{-1}$ (yellow) and the case II with $k_{p}=2\times10^{6}{\rm Mpc}^{-1}$ (red). Also $A_{p}=10^{-3}$ and $\sigma_{p}=1$ are commonly assumed. {\color{black} We showed the various constraints on $P_{\zeta}(k)$ based on PBH abundance, Pulsar timing array (PTA) and $\mu$-distortion of CMB energy spectrum with the cyan, blue and purple shaded regions respectively.} {\bf Bottom:} The fraction of DM in PBH which was formed during the EMD era and resulted from $P_{\zeta}(k)$ shown in the left panel. The assumed example $m_{\sigma}$ and $T_{\rm rh2}$ is specified in the top with the green color. {\color{black} The light blue shaded region is the constraint on $f_{\rm PBH}$ based on the microlensing.}}
  \vspace*{-1.5mm}
\label{fig:Pzetafpbh}
\end{figure*}

The fraction of PBH with a mass $M$ in DM ($f_{\rm PBH}(M)=\rho_{\rm PBH,0}/\rho_{\rm DM,0}$) can be read from the mass function which reads
\beqs
\psi(M)&\simeq&\frac{1}{M}\frac{\rho_{\rm PBH,0}(M)}{\rho_{\rm DM,0}}\cr\cr
&=&\frac{1}{M}\frac{\rho_{\rm PBH,0}(M)}{\rho_{\rm m,0}}\frac{\rho_{\rm m,0}}{\rho_{\rm DM,0}}\cr\cr
&=&\frac{1}{M}\frac{\rho_{\rm PBH,eq}(M)}{\rho_{\rm m,eq}}\frac{\Omega_{\rm m,0}h^{2}}{\Omega_{\rm DM,0}h^{2}}\cr\cr
&=&\frac{1}{M}\frac{\rho_{\rm PBH,eq}(M)}{\rho_{\rm rad,eq}}\frac{\Omega_{\rm m,0}h^{2}}{\Omega_{\rm DM,0}h^{2}}\cr\cr
&=&\frac{\Omega_{\rm m,0}h^{2}}{\Omega_{\rm DM,0}h^{2}}\frac{\beta_{\rm MD}(M)}{M}\frac{R_{\rm eq}}{R_{\rm rh2}}\cr\cr
&=&4.88\times10^{27}\times\frac{\beta_{\rm MD}(M)}{M}\sqrt{\frac{\Gamma_{\sigma}}{M_{P}}}\,,
\label{eq:psiM}
\eeqs

where $R_{\rm eq}/R_{\rm rh2}=\sqrt{H_{\rm rh2}/H_{\rm eq}}$, $H_{\rm rh2}=\Gamma_{\sigma}$, $\Omega_{\rm m,0}=0.315$, $\Omega_{\rm DM,0}=0.265$~\cite{Planck:2018vyg}, $g_{*}(R_{\rm eq})=3.36$ and $T_{\rm eq}\simeq0.75{\rm eV}$ were used for the last equality with $R_{\rm eq}$ ($R_{\rm rh2}$) the scale factor at the matter-radiation equality (the time of the moduli decay). In obtaining the third and fifth equalities, the scaling behaviors $\rho_{\rm PBH}\propto R^{-3}$ and $\rho_{\rm rad}\propto R^{-4}$ were used for PBH and radiation. 

In combination with $\beta_{\rm MD}\simeq0.056\sigma^{5}$ and Eq.~(\ref{eq:saxiondecayrate}), Eq.~(\ref{eq:psiM}) leads on to
\beq
f_{\rm PBH}(M)\simeq(2.35\times10^{5})\times P_{\zeta}(k)^{5/2}\times\left(\frac{m_{\sigma}}{100{\rm TeV}}\right)^{3/2}\,.
\label{eq:fPBH}
\eeq
where $M$-dependence of $f_{\rm PBH}$ is realized via $k$ dependence of $P_{\zeta}(k)$ on the RHS and the approximate relation $\sigma(M)\approx2(1+w)(5+3w)^{-1}\sqrt{P_{\zeta}(k)}$ was used with $w=0$. The mapping between $k$ and $M$ can be found in Eq.~(\ref{eq:MPBH}).\footnote{For $w=0$, $\sigma(M)$ is related to $P_{\zeta}(k)$ via
\beq
\sigma(M)^{2}=\left(\frac{2}{5}\right)^{2}\int{\rm d}\ln k(kr)^{4}W(k,r)^{2}P_{\zeta}\,,
\eeq
where $r$ is the horizon scale at the time of PBH formation and $W(k,r)$ is a window function. For the choice of a Gaussian window function~\cite{Young:2019osy}, we see that $\sigma(M)\approx2(1+w)(5+3w)^{-1}\sqrt{P_{\zeta}(k)}$. So in this work, assuming the Gaussian window function, we use the approximation.} In the scenario of our interest, DM population consists of axion and PBH, and thus the sum of the relic abundances of the two amounts to that of DM, i.e.
\begin{equation}
\Omega_{\text{DM}} = \Omega_a(m_{\sigma},F_a) + \Omega_{\text{PBH}}(m_{\sigma},F_a,P_{\zeta})\,.
\label{eq:DMmixed}
\end{equation} 
where
\begin{equation}
\Omega_{\text{PBH}}=\Omega_{\text{DM}} \times \int \text{d}\, \text{ln}M\, f_{\text{PBH}}(M)\,,
\end{equation}

In the top panel of Fig.~\ref{fig:Pzetafpbh}, we first show the exemplary primordial power spectrum assumed for the case I (yellow dashed) and II (red dashed). The region above the cyan, blue and purple lines are excluded by constraints from PBH abundance, Pulsar timing array (PTA) and $\mu$-distortion of CMB energy spectrum~\cite{Gow:2020bzo} respectively. In conversion of the constraints on the associated observational quantities to $P_{\zeta}(k)$, $\sigma_{p}=1$ was assumed in Eq.~(\ref{eq:Pzeta}). One can safely assume $P_{\zeta}(k)$ as large as $\mathcal{O}(10^{-3})$ for $k>10^{5}{\rm Mpc}^{-1}$ to achieve $f_{\rm PBH}$ as desired in the presence of the EMD era. For the case I and II specified in the introduction, we take the exemplary lognormal power spectrum on small scales with $A_{p}=10^{-3}$ and $\sigma_{p}=1$ for both, and $k_{p}=10^{8.5}{\rm Mpc}^{-1}$ (case I) and $2\times10^{6}{\rm Mpc}^{-1}$ (case II).\footnote{As will be seen in Sec.~\ref{sec:exp}, the PBH mass ensuring the interesting encounter rate between the tidal stream and the earth is $M_{\rm PBH}=\mathcal{O}(10^{-16})M_{\odot}-\mathcal{O}(10^{-14})M_{\odot}$. For having such a PBH, we may assume $P_{\zeta}(k)$ centered on $k=\mathcal{O}(10^{11}){\rm Mpc}^{-1}$ in the case I.}  

{\color{black} Particularly, the constraint on $P_{\zeta}(k)$ from PTA is based on NANOGrav 11 year data. We note that the future sensitivity of measurement of $\Omega_{\rm GW}h^{2}$ by SKA can be several orders of magnitude better than NANOGrav, which allows for SKA to probe $P_{\zeta}(k)=\mathcal{O}(10^{-5})$. Interestingly, as will be discussed in Sec.~\ref{sec:caseII}, the same radio telescope SKA can detect the radio signals from the encounter between UCMH and neutron star. Therefore, we expect that probing the stochastic GW background induced by the enhanced $P_{\zeta}(k)$ we assumed in the case II by SKA can be the complementary way of testing our scenario to the radio signal search by SKA. Null observation of either the stochastic GW background at the frequency range corresponding to the case II or radio signal by SKA will potentially exclude or support the case II.}

As a result of the assumed $P_{\zeta}(k)$'s in the top panel, there arise PBHs during the EMD era contributing to the DM energy density by $f_{\rm PBH}$ in Eq.~(\ref{eq:fPBH}) as shown in the bottom panel of Fig.~\ref{fig:Pzetafpbh}. The $k$-modes re-entering the horizon at $t_{*}$ and at $t_{\rm max}$ determine the minimum ($M_{\rm min}$) and maximum ($M_{\rm max}$) PBH masses formed during the EMD era. $t_{\rm max}$ is defined to be the horizon re-entry time of $k_{\rm max}$ of which associated fluctuation grows to 1 at $t_{\rm rh2}$. $k_{\rm min}$ and $k_{\rm max}$ are computed in Eq.~(\ref{eq:kmin2}) and (\ref{eq:kmax}), providing $M_{\rm min}$ and $M_{\rm max}$ when plugged in Eq.~(\ref{eq:MPBH}). For instance, for $m_{\sigma}=100{\rm TeV}$ and $T_{\rm rh2}=16{\rm MeV}$, we find $M_{\rm max}\simeq3\times10^{-3}M_{\odot}$ and $0.3M_{\odot}$ for the case I and II, respectively.

For the case II, $k_{p}\simeq k_{\rm max}$ is observed so that the peak in $f_{\rm PBH}$ contour is cut near $M_{\rm max}$ as far as PBH generated during the EMD era is concerned. In contrast, for the case I, since $k_{\rm max}\simeq10^{7}{\rm Mpc}$ is more than one order of magnitude smaller than $k_{p}=10^{8.5}{\rm Mpc}^{-1}$, the cut near the peak does not take place. Note that for $\sigma\lesssim0.005$, $\beta_{\rm MD}\simeq0.056\sigma^{5}$ does not hold because $\beta_{\rm MD}$ needs modified to encode the effect of the collapsing region's angular momentum (spin of PBH)~\cite{Harada:2017fjm}. The bottom panel of Fig.~\ref{fig:Pzetafpbh} reflects the region of $k$ satisfying $\sigma>0.005$.\footnote{Since $f_{\rm PBH}$ corresponding to $k$-space with $\sigma\lesssim0.005$ is negligibly small, it suffices for our purpose to focus on $k$-space in the top panel of Fig.~\ref{fig:Pzetafpbh} satisfying $\sigma>0.005$.}

For the assumed $P_{\zeta}(k)$, we have seen that $f_{\rm PBH}\simeq10^{-2}$ can be indeed achieved for the PBH mass ranges that each case is targeting via Fig.~\ref{fig:Pzetafpbh}. Thus the relatively weak current constraints on $P_{\zeta}(k)$ on small scales and UV-physics motivated EMD era can be interesting points to motivate one to think about $f_{\rm PBH}\lesssim\mathcal{O}(10^{-2})$ for $M_{\rm PBH}$ specified in Fig.~\ref{fig:Pzetafpbh}.\footnote{Also shown in the bottom panel as the purple shaded region is the microlensing constraint on $f_{\rm PBH}$~\cite{Macho:2000nvd,EROS-2:2006ryy,Griest:2013aaa,Oguri:2017ock,Niikura:2019kqi,Croon:2020ouk}. This microlensing constraint is relatively weak as compared to constraints in other $M_{\rm PBH}$ regimes such as $M_{\rm PBH}\gtrsim1M_{\odot}$.} Bearing in mind this possibility for PBH as a minor component of DM and the underlying physics for PBH formation in the early universe, in the coming sections we study interesting astrophysical phenomenology caused by PBHs with $f_{\rm PBH}=\mathcal{O}(10^{-2})$. Particularly since the axion is the major component of DM in our scenario, our study concerns physics resulting from the interplay between the axion and PBH. 
\vspace{2 cm}
\section{UCMH formation}
\label{sec:UCMH2}
The PBH formation via an EMD era could potentially play an important role in axion searches. A dark halo is predicted to form around an isolated and stationary PBH during the late time matter domination era via secondary infall accretion. Such an accretion leads to the formation of the so-called ultracompact minihalos (UCMHs) as shown in ~\cite{1985ApJS...58...39B, Mack:2006gz}. The growth in the mass and the radius of UCMH in time (parametrized by the redshift $z$) can be seen from ~\cite{Mack:2006gz, Berezinsky:2013fxa, Ricotti:2009bs}
\begin{align}
M_{\text{UCMH}}(z) &= 3\left(\frac{1000}{1+z} \right)\,M_{\text{PBH}}\,,\label{mhalo}\\
R_{\text{UCMH}}(z) &= 0.019\, \text{pc} \left( \frac{1000}{z+1} \right) \left( \frac{M_{\text{UCMH}}(z)}{M_{\odot}}\right)^{1/3}\,.
\label{rhalo}
\end{align}
The growth eventually leads to the formation of the steep radial density profile of the following form~\cite{Ricotti:2009bs, Bringmann:2011ut}
\begin{equation}
\begin{adjustbox}{max width=218pt}
$
\rho_{\text{UCMH}}(r)\approx0.23\,M_{\odot}\text{pc}^{-3} \left(\frac{R_{\text{UCMH}}}{r} \right)^{9/4} \left( \frac{10^2\,M_{\text{PBH}}}{M_{\text{UCMH}}} \right)^3 \,,
\label{infall} 
$
\end{adjustbox}
\end{equation}
under the approximation $\Omega_{\text{DM}}/\Omega_m\approx 1$. 

This profile indeed was confirmed by the first N-body simulation of the universe composed of a smooth DM particle background plus a small fraction of PBHs~\cite{Adamek:2019gns}. The growth of UCMHs stops when they begin to interact with non-linear structures at around $(z \sim 30-10)$~\cite{Berezinsky:2013fxa}, so that the UCMH abundance is given by $f_{\text{UCMH}}\sim 10^2 f_{\text{PBH}}$, where $f_{\text{PBH}}$ is the original fraction of DM in naked PBHs.
 The above equations only holds for $f_{\text{PBH}} \lesssim10^{-2}$ to ensure the isolated assumption.

The most inner region of the UCMH is softened by angular momentum conservation as was pointed out in~\cite{Bringmann:2011ut} and previous works take a conservative approach to model the inner region with a cut-off  within which one takes the UCMH density to be nearly constant (see, for example, ~\cite{2012PhRvD..86d3519L})~\footnote{This analysis is extremely important when the dark halo is composed of self-annihilating particles such as WIMPs. Due to DM self-annihilation mostly occurs within the most inner shells of UCMHs, constraints on the current fraction  of DM in UCMHs (and subsequently PBHs) heavily depends on the UMCH most inner density profile~\cite{Hertzberg:2020kpm, Boucenna:2017ghj, Adamek:2019gns}.}. Here we use the simplified version for UCMH density given in Eq.~(\ref{rhalo}) (for completeness, we have added in Appendix~\ref{UCMHaccurate} the density profile including an inner core).

\section{Axion Direct Detection via Tidal Streams (CASE I)}
\label{sec:exp}
In this section, we consider $f_{\rm PBH}=\mathcal{O}(10^{-2})$ with $M_{\rm PBH}\lesssim\mathcal{O}(10^{-6})M_{\odot}$, which corresponds to the case I. The formation of tidal streams from disruption of axion miniclusters during their encounters with stars in the local neighborhood as well as its implication for axion direct searches was discussed in~\cite{Tinyakov:2015cgg}. The key idea is that such tidal streams may still hold densities larger than the average. In the context of axion searches, the stream-crossing events would hold a reasonable rate of about  $1/20\, \text{yr}^{-1}$, leading to a signal amplification in axion detectors by a factor $\sim 10$ during 2-3 days.

In this section, we apply the above idea to UCMHs that possibly get generated for the case I in Fig.~\ref{fig:Pzetafpbh} (right panel). It is well known that UCMHs seeded by solar mass PBHs are extremely resistant against disruption from high speed encounters with stars. The reason is the following. The critical impact parameter, i.e. the impact parameter between the star and the UCMH which leads to a total UCMH disruption (or at least a significant loss of
mass), is much shorter than the typical UCMH radius. Only
encounters with very small impact parameters in terms of the
UCMH radius will lead to one-off disruption. Since such encounters are statistically disfavored, the disruption of UCMHs due to encounters with stars
at the solar neighborhood is negligible. However, when it comes to PBH seeds with much lighter masses, the situation changes. 

The critical impact parameter $b_c$ is defined as $\Delta E(b_c) = E_b$ with $\Delta E$ the UCMH internal energy obtained after the encounter with a star and $|E_b|$ its binding energy.  For impact parameters satisfying $b\lesssim R_{\rm UCMH}$, the distant-tide approximation is no longer valid. Thus, it becomes necessary to use the following parametrization performed in~\cite{1999ApJ...516..195C} which is valid in both $b \ll R_{\text{UCMH}}$ and $b \gg R_{\text{UCMH}}$ regimes~\cite{Hertzberg:2019exb}
\begin{align}
&\Delta E_{\gg \ll} \approx  \frac{16 \pi}{3}\left( \frac{G_N M_{\star}}{v_{\textrm{rel}} b^2} \right)^2 \times\nonumber\\
&\int_{0}^{R_{\textrm{UCMH}}} dr\, r^4 \rho_{\textrm{UCMH}}(r) \left( 1 + \frac{4 r^4}{9 b^4}  \right)\left( 1+ \frac{2r^2}{3b^2}  \right)^{-4}\,,\label{eq:full}\\
& \approx6.5 \times 10^{-21}\,M_{\odot} \left( \frac{M_{\star}}{M_{\odot}} \right)^2 \left( \frac{220\,\textrm{km/s}}{v_{\textrm{rel}}} \right)^2\left(  \frac{R_{\textrm{UCMH}}}{b} \right)^2\times  \nonumber\\
& \left( \frac{M_{\textrm{PBH}}}{M_{\odot}} \right)^{1/3} \left( \frac{10^2\,M_{\textrm{PBH}}}{M_{\textrm{UCMH}}}  \right)^{5/3}\left[ \frac{S(R_{\textrm{UCMH}}/b)}{S(1)} \right]\,,\label{Esoso}
\end{align} 
where $M_{\star}$ is the mass of the encountered star, $S(1)=0.066$ and
\begin{equation}
S(R_{\textrm{UCMH}}/b) = \sum_{i=1}^{4} C_{i}\, {}_{2}F_{1}  \left[\frac{3}{8},i,\frac{11}{8},-\frac{2}{3}\left( \frac{R_{\textrm{UCMH}}}{b}\right)^2\right]\,,\nonumber
\end{equation}
with $\left\{C_i\right\}_{i=1}^{4} = (1,-3,4,-2)$ and $_2 F_1$ is the known Gauss hypergeometric function. 

On the other hand, the UCMH binding energy is given by
\begin{equation}
E_b \approx 2.4 \times 10^{-10}\,M_{\odot} \left(  \frac{M_{\textrm{UCMH}}}{10^2\,M_{\textrm{PBH}}} \right)^{2/3}\left( \frac{M_{\textrm{PBH}}}{M_{\odot}} \right)^{5/3}\,.\label{Eb}
\end{equation}
The critical impact parameter can then be obtained by solving numerically $\Delta E(b_c) = E_b$ with the use of Eqs.~(\ref{Esoso}) and (\ref{Eb}). 

In the regime $b \ll R_{\text{UCMH}}$, we can expand $S(R_{\textrm{UCMH}}/b)$ of Eq.~(\ref{Esoso}) in a Taylor series at around $b/R_{\textrm{UCMH}}=0$ to obtain~\cite{Hertzberg:2019exb}
\begin{align}
&\frac{b_c}{R_{\textrm{UCMH}}} \approx 1.5\times 10^{-8} \left( \frac{M_{\star}}{M_{\odot}} \right)^{8/5}\times \nonumber \\
&
 \left( \frac{220\,\frac{km}{s}}{v_{\textrm{rel}}} \right)^{8/5} \left( \frac{M_{\odot}}{M_{\textrm{PBH}}} \right)^{16/15}
 \left( \frac{10^2\, M_{\textrm{PBH}}}{M_{\textrm{UCMH}}} \right)^{28/15}\,.\label{bc}
\end{align}

In the other regime $b \gg R_{\text{UCMH}}$, we can expand Eq.~(\ref{eq:full}) around $(r/b) =0$. We see that the expression for the gained internal energy asymptotically approaches the well-known distant-tide approximation when $b/R_{\textrm{halo}} \rightarrow \infty$~\cite{2008gady.book.....B, 1958ApJ...127...17S}, i.e.
\begin{equation}
\Delta E_{\gg} \approx \frac{4 \langle r^2 \rangle}{3}\frac{G_N^2 M_{\star}^2M_{\textrm{UCMH}}}{v_{\textrm{rel}}^2b^4}\,,\label{eq:distant-tide}
\end{equation}
where
\begin{equation}
 \langle r^2 \rangle = \left(\frac{1}{M_{\textrm{UCMH}}}\right)\int^{R_{\text{UCMH}}} d^3r\, r^2 \rho_{\textrm{UCMH}}(r)
 \end{equation}
 is the mass-weighted mean-square UCMH radius.

\begin{figure}[t!]
\centering
\hspace*{-5mm}
\includegraphics[width=0.425\textwidth]{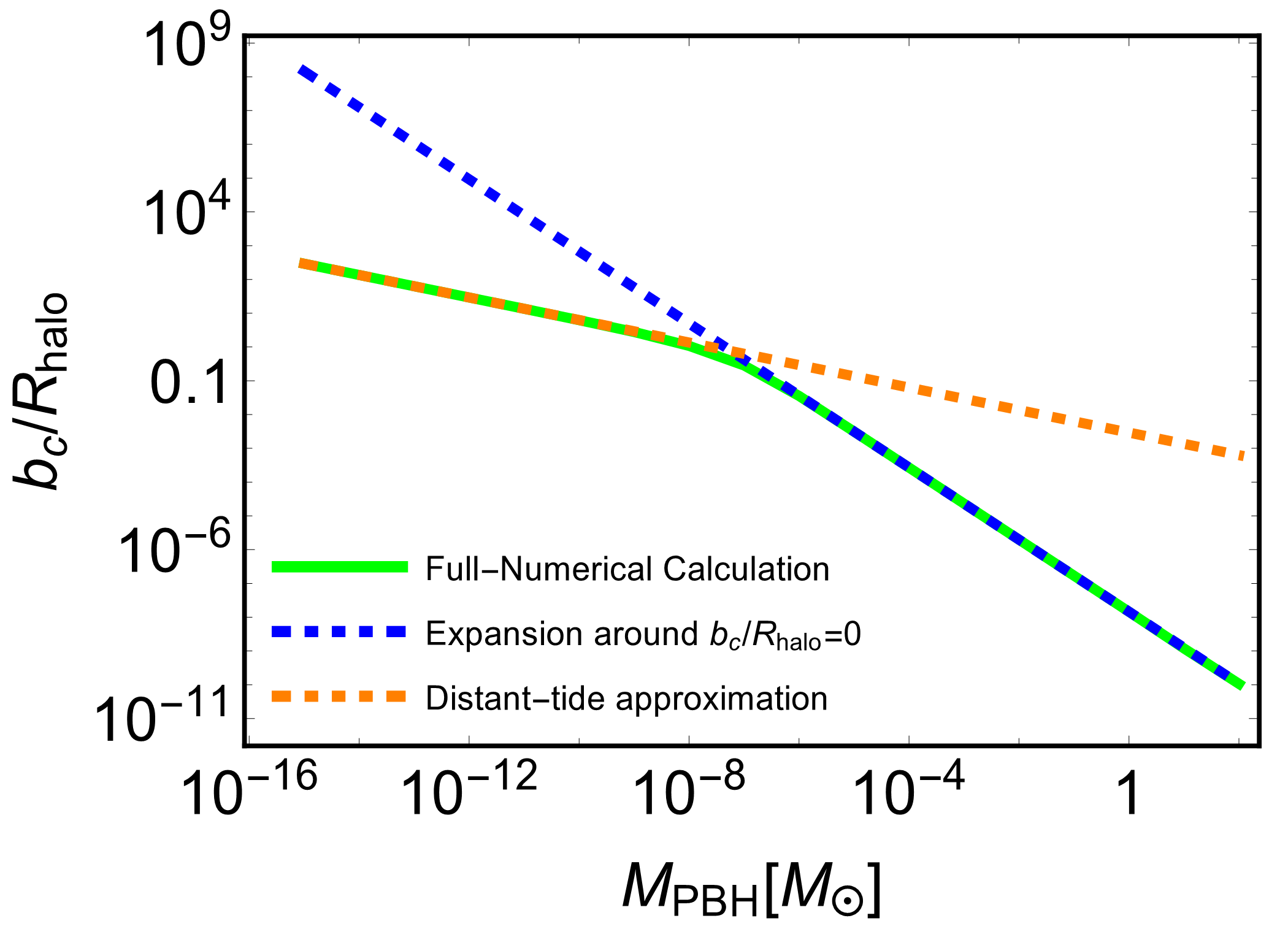}
\caption{Critical impact parameter normalized by the UCMH radius using a full numerical calculation (solid green line), an expansion around $b_c/R_{\text{UCMH}}=0$ (dashed blue line), and the distant-tide approximation (dashed orange line). We have fixed $M_{\text{UCMH}}=10^2\,M_{\text{PBH}}$ and $v_{\text{rel}}=220\,\text{km/s}$.} 
\vspace*{-1.5mm}
\label{fig:UCMHbc}
\end{figure}

In Fig.~\ref{fig:UCMHbc}, we show the result of the full numerical calculation for the critical impact parameter (solid green line) together with the approximation given in Eq.~(\ref{bc}) valid for $b\ll R_{\text{UCMH}}$ (dashed blue line) and the distant-tide approximation valid for $b\gg R_{\text{UCMH}}$ (dashed orange line). We see that there is a clear break in the curve at around $M_{\text{PBH}} \sim \mathcal{O}(10^{-8})\,M_{\odot}$, which separates the regimes in which the critical impact parameter is larger or smaller than the UCMH radius. For $M_{\text{PBH}} \lesssim 10^{-8}\,M_{\odot}$, the critical impact parameter is well-described under the distant-tide approximation.

Consider a star field with a density $n_{\star}$ and suppose that the minicluster spends a time $t_{\text{cross}}$ within such a field. With account taken of the sum of one-off disruptive events with $b < b_c$ and the contribution from multiple encounters with $b \geq b_c$, the UCMH disruption probability during one-single crossing ($P_s$) through the galactic disk reads~\cite{Schneider:2010jr},\footnote{For each encounter, the disruption probability is quantified as $\Delta E(b)/E_{b}$. For $b < b_{c}$, $\Delta E(b)=E_{b}$ was used.}
\begin{align}
 P_{s} &= 2\pi n_{\star} v_{\text{rel}} t_{\text{cross}} \times \left( \int_0^{b_c} b db +  \frac{1}{E_b}\int_{b_c}^{\infty} \Delta E(b)bdb\right)\,,\\
 &= \left( \frac{\pi b_c^2}{M_{\star}}\right) \times S_{\perp} + \frac{2\pi}{M_{\star}} \left(\frac{1}{E_b}\int_{b_c}^{\infty} \Delta E(b)bdb\right)\times S_{\perp}\,,\label{eq:fullP}
\end{align}
where we took the field as a column of stars with density $n_{\star}$ and height $H$, so that the (perpendicular) crossing time is $t_{\text{cross}} = H/v_{\text{rel}}$ and $n_{\star} = S_{\perp}/(H M_{\star})$ with $S_{\perp} \approx 35\,M_{\odot}\text{pc}^{-2}$~\cite{1989MNRAS.239..571K} the perpendicular surface mass density in the local neighborhood.\footnote{The mass dependence of number density of stars becomes suppressed for $M_{\star}\gtrsim 0.5M_{\odot}$ so that we choose $M_{\star}=1M_{\odot}$ as a characteristic value~\cite{10.1111/j.1365-2966.2007.11397.x, Kroupa:2002ky}.} 
\textcolor{black}{By assuming isotropically distributed trajectories and setting the maximum trajectory length to be no longer than $\mathcal{O}(10)$ kpc to avoid a logarithmic divergence in the angular integration, authors in~\cite{Tinyakov:2015cgg} shows that the probability $P_s$ in Eq.\,(\ref{eq:fullP}) needs to be re-scaled as $P_s \rightarrow 4 P_s$ to include different directions.}

To estimate the total probability for UCMH disruption, $P_{\text{Total}}$, we multiply the disruption probability during a single disk crossing by the total number of crossings during the age of the Milky Way $T_{\text{MW}} \sim 10\,\text{Gyr}$, $N_{\text{cross}} \sim 100$. \textcolor{black}{Since $P_{\text{Total}}=P_{\text{Total}}(t)$ by construction, a total probability equal to one means that a time  $t=T_{\text{MW}}$ is needed to disrupt all UCMHs. A $P_{\text{Total}} > 1$ means that such a total disruption occurs during a total time $t < T_{\text{MW}}$.}
\begin{figure}[t!]
\centering
\hspace*{-5mm}
\includegraphics[width=0.425\textwidth]{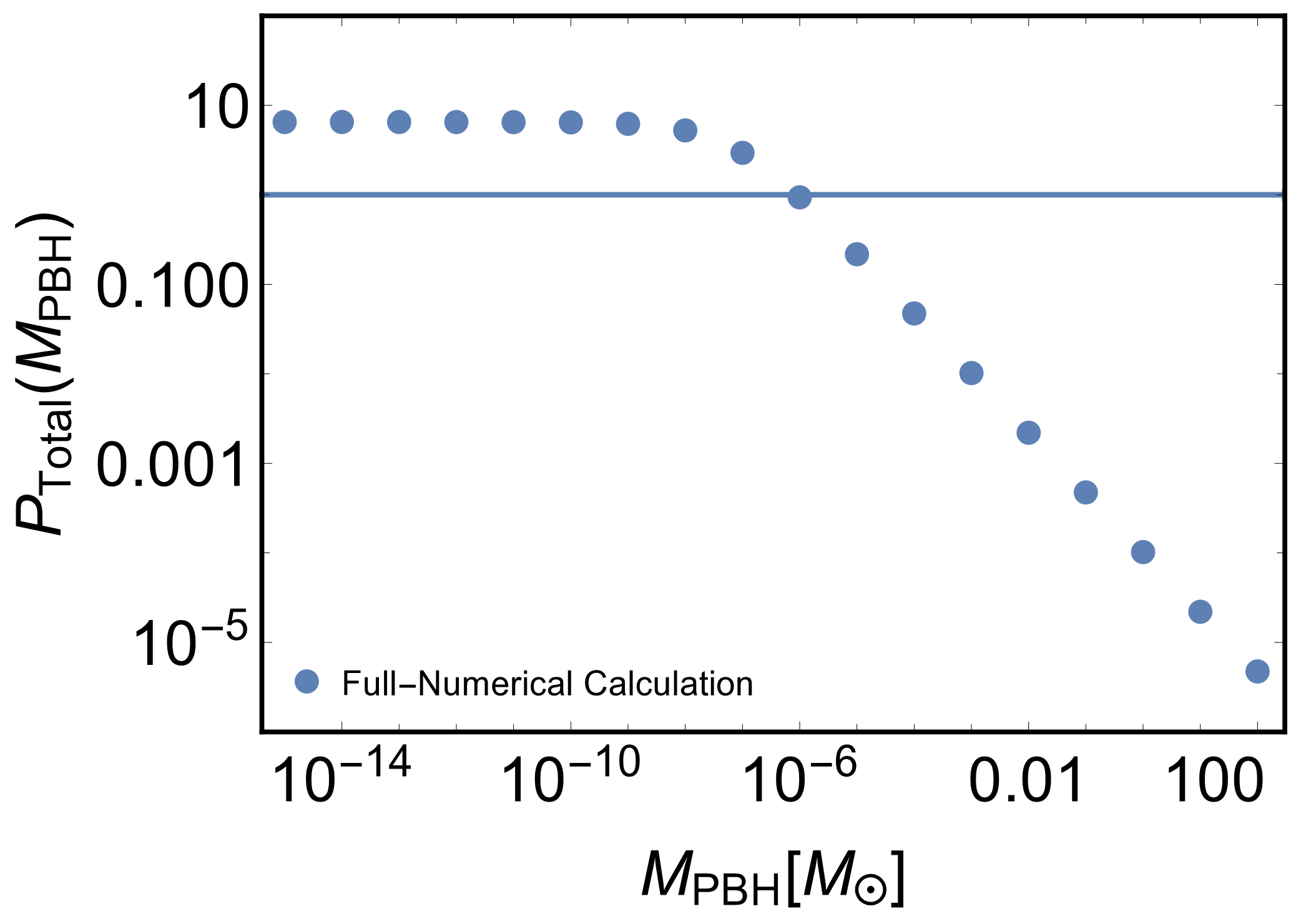}
\caption{Probability of UCMH disruption in the local neighborhood using a full numerical calculation (blue points). We have fixed $M_{\text{UCMH}}=10^2\,M_{\text{PBH}}$ and $v_{\text{rel}}=220\,\text{km/s}$. The blue line indicates $P_{\text{Total}}=1$.} 
\vspace*{-1.5mm}
\label{fig:UCMHPt}
\end{figure}

Fig.~\ref{fig:UCMHPt} shows the total disruption probability of UCMHs in the local neighborhood based on Eq.~(\ref{eq:fullP}). We have used $M_{\text{UCMH}}=10^2\,M_{\text{PBH}}$ and $v_{\text{rel}}=220\,\text{km/s}$. We see that all UCMHs
are disrupted for $M_{\text{PBH}} \lesssim 10^{-6}M_{\odot}$.
Such a phenomenon could potentially have a great impact on axion direct searches. 
 
On disruption of UCMHs, debris arise, which become tidal streams along the UCMH path. For orbits with small ellipticity, the streams can be considered to be one-dimensional structures of a length $L_{st}$ and a cross-section $\sim \mathcal{O}(R_{\text{UCMH}}^2)$. Such a tidal stream formation process was modeled in~\cite{Schneider:2010jr, Angus:2006vp} for DM clumps with a mass $\sim 10^{-6}M_{\odot}$. Here we closely follow ~\cite{Schneider:2010jr, Tinyakov:2015cgg} to estimate the typical length of tidal streams $L_{\text{st}} \sim \sigma_a \Delta t$ with $\sigma_a$ the axion velocity dispersion of the initial UCMH, and $\Delta t$ the time between UCMH disruption and today. Under this approximation, the UCMH grows in volume by a factor $\sigma_a \Delta t/R_{\text{UCMH}}$ during a time period $\Delta t$. This gives rise to dilution of the UCMH radial density in a plane perpendicular to the stream axis
\begin{align}
&\rho_{\text{st}}(r) = \rho_{\text{UCMH}}(r) \times \frac{r}{\sigma_a(r) \Delta t}\,,\,\\
& \approx 0.06\, M_{\odot} \text{pc}^{-3} \left( \frac{10^{-2}}{r/R_{\text{UCMH}}} \right)^{9/8} \left(\frac{10^2\,M_{\text{PBH}}}{M_{\text{UCMH}}} \right)^{3/2}\,,
\end{align}
where $\sigma_a(r) = (0.3\, G_N M_{\text{UCMH}}(r)/r)^{1/2}$ as obtained in Appendix~\ref{EGDF} via the corresponding ergodic distribution function of the UCMH. To obtain a conservative estimate, we have used  $\Delta t = 10\,\text{Gyr}$. Given the local DM density $\rho_{\text{local}} = 0.008 \pm 0.003\,M_{\odot}\text{pc}^{-3}$~\cite{Bovy:2012tw}, it is realized that tidal streams can reach densities about an order of magnitude larger than $\rho_{\text{local}}$ in its inner region at $r\simeq 10^{-2}R_{\text{UCMH}}$.
\begin{figure}[t!]
\centering
\hspace*{-5mm}
\includegraphics[width=0.425\textwidth]{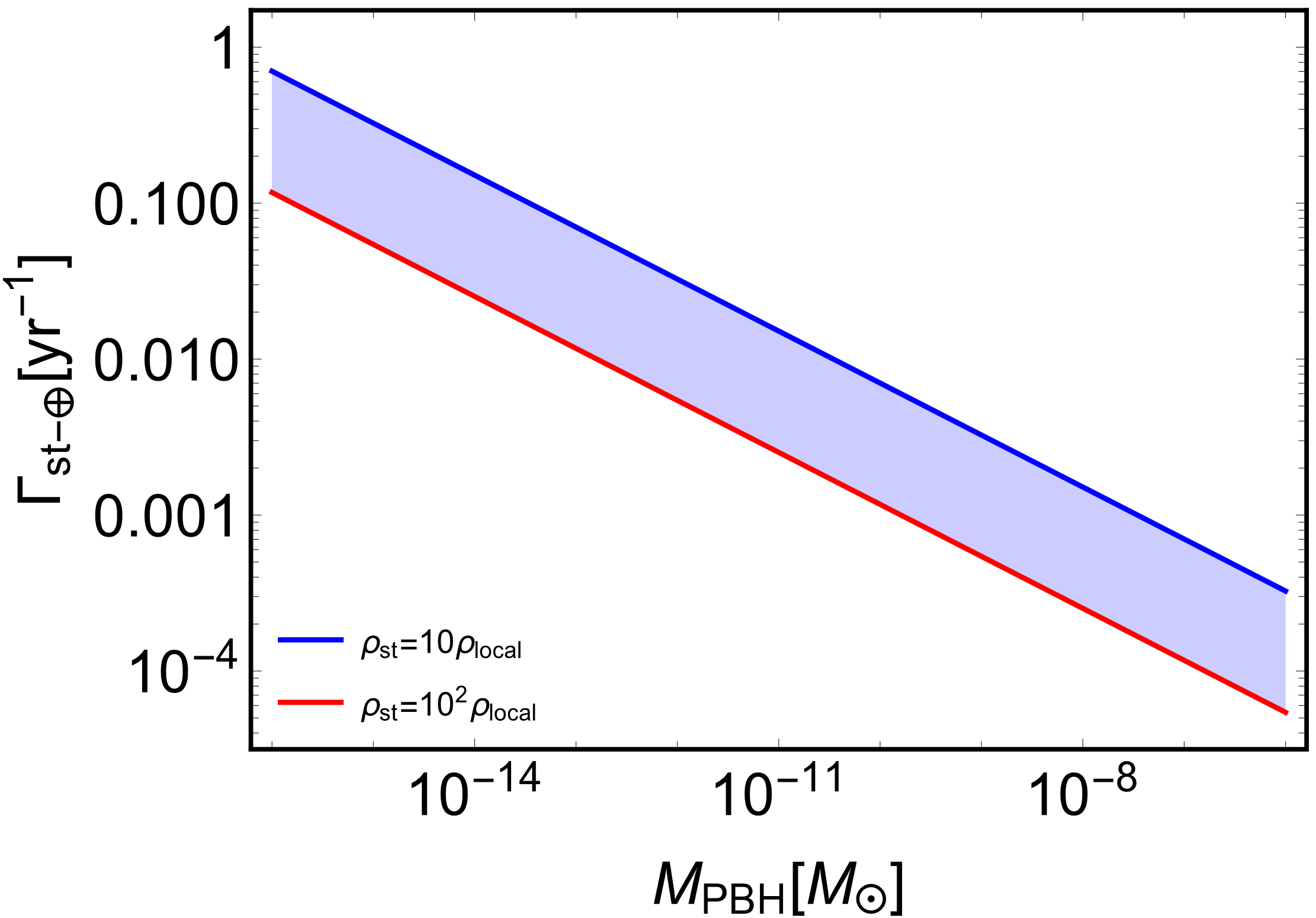}
\caption{Earth and tidal stream encounter rate for UCMHs in terms of the central PBH mass according to Eq.~(\ref{eq:gamma}). We have taken $M_{\text{UCMH}}=10^2\,M_{\text{PBH}}$ as the original UCMH mass.}
\vspace*{-1.5mm}
\label{fig:Gamma}
\end{figure}

The probability that the Earth passes by one of tidal streams is expected to be proportional to the total number of UCMH within a certain volume, namely the filling factor. Given that for $M_{\text{PBH}} \lesssim 10^{-6}\,M_{\odot}$, the local number density of tidal streams is equal to the initial PBH number densities, i.e. $n_{\text{st}} = n_{\text{PBH}}$, the filling factor (FF) reads
\begin{align}
 &\text{FF(r)} = \left( \frac{f_{\text{PBH}}\rho_{\text{local}}}{M_{\text{PBH}}} \right) \times \left( \frac{4\pi r^3}{3} \frac{\sigma_a(r) \Delta t}{r} \right)\,,\\ 
 & \approx 0.001 \left( \frac{f_{\text{PBH}}}{10^{-2}} \right) \left( \frac{r/R_{\text{UCMH}}}{10^{-2}} \right)^{15/8}\left( \frac{M_{\text{UCMH}}}{10^2\,M_{\text{PBH}}} \right)^{5/2}\,.\label{eq:prob}
\end{align}

Using Eq.~(\ref{eq:prob}), we estimate the 
Earth-tidal stream encounter rate as 
\begin{equation}
    \Gamma_{{\rm st}-\oplus}(r)= \frac{\text{FF}(r)}{\tau(r)} = \text{FF}(r) \times \frac{v_{\text{rel}}}{2 r}\,,\label{eq:gamma}
\end{equation}
where
\begin{equation}
\begin{adjustbox}{max width=218pt}
$
\tau(r) \approx 0.6\,\text{yr}\,\left( \frac{r/R_{\text{UCMH}}}{10^{-2}} \right) \left( \frac{M_{\text{UCMH}}}{10^2\,M_{\text{PBH}}} \right)^{4/3}  \left( \frac{M_{\text{PBH}}}{10^{-8}M_{\odot}} \right)^{1/3}\,.\label{eq:tcross} 
$
\end{adjustbox}
\end{equation}
Here $\tau(r)$ is the typical crossing time of the Earth through a high dense DM zone of radius $r$ within the tidal stream. We are assuming that Earth is crossing such a tidal perpendicularly to its axis. 

\textcolor{black}{Thinking of DM direct searches, we require that such a high axion zone holds a density  $\rho_{\text{st}} \in [10-100]\rho_{\text{local}}$ to calculate in Figure~\ref{fig:Gamma} the Earth-tidal stream encounter rate in terms of the central PBH mass. 
One can see} that the lighter the central PBH in (undisrupted) UCMHs is, the larger the encounter rate for fixed $\rho_{\text{st}}$ becomes. Interestingly, $\Gamma_{{\rm st}-\oplus}$ can be high enough so that the Earth can encounter the tidal stream of axion DM with the enhanced $\rho_{\text{st}}\gtrsim 10\rho_{\text{local}}$ once every (1-10) years for $M_{\text{PBH}}\lesssim 10^{-14}M_{\odot}$, which is comparable to that found in~\cite{Tinyakov:2015cgg} for the case of axion miniclusters.  
  
If the aforementioned scenario becomes the case, one can expect the significantly enhanced chance for direct search experiments on the Earth to detect axion with the decay constant $F_a \gtrsim 10^{12}\,\text{GeV}$) (see the bottom panel in Fig.~\ref{fig:EMDevolve}). We notice that the future planned ABRACADABRA~\cite{Kahn:2016aff} experiment can probe $g_{a\gamma\gamma}-m_{a}$ parameter space that the scenario of our interest indicates.

The coupling of axion to photon via the operator $\mathcal{L}=-(g_{a\gamma\gamma}/4)aF^{\mu\nu}\tilde{F}_{\mu\nu}$ modifies the Maxwell equations so that there arises an axion-sourced effective current $\vec{J}_a(t)$ in the presence of a static magnetic field background $\vec{B}_0$. Given $a=(\sqrt{2\rho_{\rm DM}}/m_{a})\sin(m_{a}t)$, the oscillating $\vec{J}_a(t)$ induces the real oscillating magnetic field with frequency $m_{a}$ of which flux through the pick-up loop reads
\begin{equation}
\Phi_a(t)=g_{a\gamma\gamma} |\vec{B}_0| \sqrt{2\bar{\rho}_a}\text{cos}(m_a t)\,,
\label{eq:flux}
\end{equation}
where $\bar{\rho}_a$ is the axion background energy density. 

In light of our scenario, we expect there can be two interesting effects on the experiment. Firstly, for the case where the experimental sensitivity is still not good enough to probe ($g_{a\gamma\gamma}$,$m_{a}$), we expect a temporary improvement of the experimental sensitivity on the earth's encountering the tidal stream with $\bar{\rho}_{\text{st}}$. For a given experimental set-up characterized by the magnitude of the external magnetic field and volume containing it, i.e. ($B_{0,{\rm max}},V_{\rm B}$), what matters in the flux in Eq.~(\ref{eq:flux}) is the product $g_{a\gamma\gamma}\sqrt{\bar{\rho}_{a}}$. This means that when there is an enhancement in $\bar{\rho}_{a}$ due to the earth's encountering the tidal stream with $\bar{\rho}_{\text{st}}$, the experimental sensitivity for $g_{a\gamma\gamma}$ increases by
\begin{equation}
g_{a\gamma\gamma} = g_{a\gamma\gamma}(\bar{\rho}_a) \left( \frac{\bar{\rho}_{\text{st}}}{ 0.008\,M_{\odot}/\text{pc}^3} \right)^{-1/2}\,.
\end{equation}
 
 Therefore, even if ($B_{0,{\rm max}},V_{\rm B}$) are not large enough and accordingly the sensitivity is not good enough to probe QCD axion parameter space with a large $F_{a}$, still there can be a chance for the temporary improved sensitivity in our scenario with $\bar{\rho}_{\rm st}=(10-100)\rho_{a}$.
 
 On the other hand, in the case where the experimental sensitivity is already sufficiently good leading to detection of the oscillating flux, the scenario predicts a temporary amplification of the flux while the encountering happens. This possibility can be a unique prediction of the scenario and serve as the crucial way to test the scenario along with efforts to search for PBHs with $M_{\rm PBH}=\mathcal{O}(10^{-15})-\mathcal{O}(10^{-7})M_{\odot}$.

\section{Axion indirect detection via transient radio signals (CASE II)}
\label{sec:caseII}
 
 In this section, we consider $f_{\rm PBH}\simeq10^{-3}$ with $M_{\rm PBH}=\mathcal{O}(1)M_{\odot}$, which corresponds to the case II.\footnote{The assumption for $f_{\rm PBH}\simeq10^{-3}$ with $M_{\rm PBH}=\mathcal{O}(1)M_{\odot}$ is translated to $f_{\rm UCMH}\simeq10^{-1}$ with $M_{\rm PBH}=\mathcal{O}(100)M_{\odot}$. We note that this assumption might be in tension with the constraint on MACHO DM fraction from the
survival of a star cluster near the core of Eridanus II and a sample of compact ultrafaint dwarfs~\cite{Brandt:2016aco}. However, even if we assume $f_{\rm PBH}=\mathcal{O}(10^{-4})$, the encounter rate in Eq.~(\ref{eq:encounterrate}) is still not too small for our purpose.} The current presence of UCMHs in the galaxy opens up a fascinating window for axion detection via the transient radio signals coming from the axion-photon conversion during UCMH-Neutron star encounters. In the highly magnetized environment found on magnetospheres of neutron stars (NSs), axion can resonantly get converted into photons via the inverse Primakoff effect when its mass matches the plasma frequency~\cite{Hook:2018iia}. Taking into account (1) tidal disruption in the Milky Way, (2) a total number of $8\times 10^8$ NSs distributed in the galactic disk and bulge, (3) a galactic model which complies with theoretical modelling and fits observational constraints~\cite{2017MNRAS.465...76M}, and (4) UCMHs undergoing circular orbits around the galactic center, the UCMH-NS total encounter rate integrated up to 100 pc is
 calculated to be~\cite{Nurmi:2021xds}
 \begin{equation}
\Gamma^{100}_{\text{UMCH-NS}} \approx 0.17\,\text{day}^{-1}\,\left(\frac{f_{\text{PBH}}}{10^{-3}} \right) \left( \frac{M_{\text{PBH}}}{M_{\odot}} \right)^{-1/3}\,.    \label{eq:encounterrate}   
 \end{equation}
The larger the central PBH mass is, the smaller the encounter rate we have because the decrease in the UCMH number density dominates over the increase in the cross section. The typical crossing time for $M_{\text{UCMH}}=10^2\,M_{\odot}$ and $M_{\text{PBH}}=M_{\odot}$ is about $\sim10^4\,\text{kyr}$ (see Fig. 7 in ~\cite{Nurmi:2021xds}). Such a long crossing time leads to a transient radio signal when the axion density at the NS conversion radius is high enough to exceed the telescope sensitivity. 
 
 In this section, we study the above scenario in the context of UCMHs seeded by solar mass PBHs, which were produced for the case II in Fig.~\ref{fig:Pzetafpbh}. In particular, we extend the analysis performed in~\cite{Nurmi:2021xds}  for QCD axion window to the case of $F_{a} \gtrsim 10^{12}~\text{GeV}$. We adopt the Goldreich and Julian (GJ) model~\cite{1969ApJ...157..869G} of the NS magnetosphere and closely follows~\cite{Hook:2018iia}. For an axion
mass in the MHz range, the axion-photon resonant conversion takes place in a small region around the conversion radius $r_c$. 

In the case of aligned or slightly oblique NSs with a radius $R_{\rm NS}$, e.g. a zero or small angle $\theta_m$ between the NS rotating axis and the magnetic field, the power radiated per unit solid angle at the observation angle $\theta$ and zeroth order in $\theta_m$ is well approximated by
\begin{equation}
\begin{adjustbox}{max width=218pt}
$
    \frac{d\mathcal{P}(\theta)}{d\Omega} \approx \frac{g_{a\gamma\gamma}^2B_{0}^2\rho_{a}(r_c)\pi v_c}{6 m_a}\left(\frac{R^2_{\text{NS}}}{r_c} \right)^3\left[3\text{cos}^2(\theta)+1 \right]\,, \label{DPOMEGA}
    $
\end{adjustbox}
\end{equation}
where $\rho_a(r_c)$ ($v_c$) is the axion density (velocity) evaluated at the conversion radius, $B_0$ is NS magnetic field strength at the poles and the WKB and stationary phase approximations have been used. The conversion radius reads
\begin{align}
&r_c(\theta) = 224\,\text{km}\left( \frac{R_{\text{NS}}}{10\,\text{km}} \right) \left(\frac{B_0}{10^{14}\,\text{G}} \right)^{1/3}  
\left(\frac{\text{sec}}{P} \right)^{1/3}\times\nonumber \\
&\left(\frac{\text{GHz}}{m_a} \right)^{2/3}|3\text{cos}^2(\theta)-1|^{1/3}\,,\label{eq:rc}
\end{align}
where $P$ is the period of NS spin. 

The power radiated per solid angle rapidly decays as $r_{c}$ becomes larger, and shows dependence on the axion mass and NS astrophysical properties. The Equation~(\ref{DPOMEGA}) only holds for $r_c > R_{\text{NS}}$ since no resonant conversion occurs within the NS. As the NS encounters the UCMH, axion particles fall into the NS with a velocity at the conversion radius  %
\begin{equation}
\begin{adjustbox}{max width=218pt}
$
  v_c \approx \left(\frac{2G_N M_{\text{NS}}}{r_c}\right)^{1/2} \approx 0.1 \text{c}  \left(\frac{M_{\text{NS}}}{M_{\odot}}  \right)^{1/2} \left( \frac{224\,\text{ km}}{r_c} \right)^{1/2}\,.\label{vc} 
  $
\end{adjustbox}
\end{equation}

Given Eq.~(\ref{DPOMEGA}), now the quantity of our interest is the spectral flux density defined by $S = (d\mathcal{P}/d\Omega)/(d^2 \Delta \nu)$ with $\Delta \nu$ and $d$ denoting the signal bandwidth and the distance from the encounter to the Earth respectively. In accordance with the above equations, the spectral flux density can be expressed in terms of
the NS mass ($M_{\rm NS}$) and radius ($R_{\rm NS}$), axion mass ($m_{a}$), and axion-photon coupling constant ($g_{a\gamma\gamma}$) as~\cite{Nurmi:2021xds}
\begin{equation}
S(\theta) = \tilde{S} \times   \frac{3\text{cos}^2(\theta)+1}{|3\text{cos}^2(\theta)-1|^{7/6}}\,,  
\end{equation}
where $\tilde S = S(\theta=\pi/2)$ and
\begin{equation}
\begin{adjustbox}{max width=224pt}
$
\tilde{S} \sim \, \mu\text{Jy}\,\left( \frac{\rho_{a}(r_c)}{0.23\,M_{\odot}\text{pc}^3} \right) \,
\left( \frac{P}{1\,\text{s}} \right)^{7/6}\left( \frac{B_0}{10^{14}\,\text{G}} \right)^{5/6} \times\nonumber   
$
\end{adjustbox}
\end{equation}
\vspace{-0.4cm}
\begin{equation}
\begin{adjustbox}{max width=220pt}
$
\left( \frac{g_{a\gamma\gamma}}{10^{-12}\,\text{GeV}^{-1}}\right)^2\left( \frac{R_{\text{NS}}}{10\,\text{km}} \right)^{5/2} \left(\frac{M_{\text{NS}}}{M_{\odot}}  \right)^{1/2} \left( \frac{m_a}{\text{GHz}} \right)^{4/3}\times\nonumber   
$
\end{adjustbox}
\end{equation}
\vspace{-0.4cm}
\begin{equation}
\begin{adjustbox}{max width=220pt}
$
\left( \frac{100\,\text{pc}}{d}\right)^{2} \left( \frac{1\,\text{kHz}}{\Delta \nu} \right)\,.\hspace{3.6 cm}\label{eq:S}
$
\end{adjustbox}
\end{equation}

For the regime deep inside UCMH, the axion density within UCMH reaches very high values several orders larger than the local DM density as shown in Eq.~(\ref{infall}). 
For $M_{\text{UCMH}} = 10^2 M_{\text{PBH}}$, the gravitational potential within the UCMH is dominated by the central PBH at radii $r \lesssim 2\times 10^{-3}R_{\text{UCMH}}$. In such a scenario, the estimation of the axion DM speed distribution is given by Eq.~(\ref{eq:fMPBH}) in Appendix~\ref{EGDF}. Following~\cite{Edwards:2019tzf}, we apply the Liouville's theorem~\cite{Liouville:1838zza} to find the axion density at the conversion radius as\footnote{Given the relation from the Liouville's theorem $\rho_{a}^{r_{c}}f_{r_{c}}(\vec{v})=\rho_{a}^{\infty}f_{\infty}(\vec{v}_{\infty}(\vec{v}))$, we integrate over $v_{c}$ to obtain Eq.~(\ref{eq:rhorc}). After the encounter, $\vec{v}_{\infty}(\vec{v})$ is mapped to $\vec{v}$.} 
\begin{align}
  &\rho_{a}(r_c)=\,\rho_{\text{R}}\left(\frac{R_{\text{UCMH}}}{G_N M_{\text{PBH}}}\right)^{9/4}\,\frac{45\Gamma[5/4]}{4\sqrt{8\pi}\Gamma[7/4]}\times\nonumber\\
  &\int_{v_{a,\text{min}}}^{v_{a,\text{max}}}v_a^2\left(\Psi_{\text{PBH}}+\Psi_{\text{NS}}-\frac{v_a^2}{2}\right)^{3/4}dv_a\,,\label{eq:rhorc}  
\end{align}
where $\rho_R = \rho_{\text{UCMH}}(R_{\text{UCMH}})$ in Eq.~(\ref{infall}),  $v_{a,\text{min}} = \sqrt{2\Psi_{\text{NS}}}$,  $v_{a,\text{max}} = \sqrt{2(\Psi_{\text{NS}}+\Psi_{\text{PBH}})}$, with $\Psi_{\text{NS}} = G_N M_{\text{NS}}/r_c$ and $\Psi_{\text{PBH}}(r) = G_N M_{\text{PBH}}/r$.  

The signal is peaked around the axion mass, i.e. $\nu_{\text{peak}} = m_a/(2\pi)\approx24\,\text{GHz}\,(10^{-4}\,\text{eV}/m_{a})^{-1}$.
For the case of our interest where there exists an EMD era followed by entropy production, an axion decay constant lying in the range $F_a \sim (10^{12}-10^{16})\,\text{GeV}$ leads to an axion DM (see bottom panel in Fig.~\ref{fig:EMDevolve}). Given $m_a(0) = (78\,\text{MeV})^2/F_a$, this axion decay constant range is associated with a peak frequency $\nu_{\text{peak}} \sim (0.1\, \text{MHz}-1\,\text{GHz})$. This frequency range is split in two different regimes by the critical frequency $\nu_{\text{crit}} \sim 30\,\text{MHz}$. While for $10^{12}\,\text{GeV} \lesssim F_a \lesssim 5\times 10^{13}\,\text{GeV}$ the radio signal could be in principle detectable on the Earth, larger axion decay constants would require lunar or space-based facilities. 

The absorption and scattering of low frequency photons by the ionosphere makes highly challenging the detection of frequencies less than the critical one. Indeed, the fact that
the ionosphere plasma frequency is around 15 MHz (10 MHz) on the day (night) side of the Earth near sunspot maximum (minimum) makes this layer opaque to all lower frequencies.  The Orbiting Low Frequency Antennas for
Radio Astronomy Mission (OLFAR), which plans to put in orbit thousands of  nano satellites on the far side of the moon, and other similar projects, will
offer in the future a chance for detection of such axions with $5\times10^{13}\,\text{GeV} \lesssim F_a \lesssim10^{16}\,\text{GeV}$ as discussed in~\cite{BENTUM2020856, 2016ExA....41..271R}. 
Here we will focus on the regime  $10^{12}\,\text{GeV} \lesssim F_a \lesssim 5\times 10^{13}\,\text{GeV}$ and leave space-based analysis of the proposed scenario for future work.

For aligned NSs, the signal bandwidth is proportional to the initial axion dispersion within UCMH according to $\Delta \nu \sim \nu_{\text{peak}} \sigma_a^2$~\cite{Hook:2018iia}, where we have $\sigma_a = (0.3\, G_N M_{\text{UCMH}}/R_{\text{UCMH}})^{1/2} \sim 10^{-6}$,  Eq.~(\ref{eq:sigmaa}). For the case of misaligned NSs, there exists an additional contribution which, in our case, largely dominates the signal broadening. The temporal dependence from the co-rotation of the plasma with the NS modifies the frequencies generating a
Doppler broadened signal with a width varying in time.
The shape of the signal as well as its time dependence is fully calculated in~\cite{Battye:2021xvt} via ray tracing 
and the use of Hamiltonian optics for a dispersive medium in curved spacetime. Here, we are mostly interested in an order of magnitude estimate of the Doppler broadening. 

The characteristic size of the ray frequency shift with respect to the unperturbed frequency ($\nu_{\text{peak}} \simeq m_a$) to leading order in $\Omega = 2\pi/P$ is given by (see Eq.~(43) in~\cite{Battye:2021xvt}) 
\begin{equation}
\Delta \nu \sim m_a^{1/3} \Omega^{4/3} R_{\text{NS}} \text{sin}\,\theta_m\, \left( \frac{2\pi\alpha_{\text{EM}} B_0}{e m_e} \right)^{1/3}\,,\textcolor{white}{xxx}
\end{equation}
\vspace{-0.3 cm}
\begin{equation}
\begin{adjustbox}{max width=218pt}
$
\sim 4\,\text{MHz}\,\left( \frac{m_a}{\text{GHz}} \right)^{1/3} \left(  \frac{\text{s}}{P}\right)^{4/3} \left( \frac{B_0}{10^{14}\,\text{G}} \right)^{1/3} \left(\frac{R_{\text{NS}}}{10\,\text{km}} \right)\,\text{sin}\,\theta_m\,.\label{eq:B} 
$
\end{adjustbox}
\end{equation}
This estimate for the frequency shift agrees very well with a previous estimates performed in~\cite{Battye:2019aco}. The shift increases with increase in the axion mass. 

The spectral flux density estimated in Eq.~(\ref{eq:S}) only holds for sources whose signal bandwidth $\Delta \nu$ is wider than the intrinsic frequency resolution of the telescope~\cite{Hook:2018iia} and for conversion radii larger than the NS radius, i.e. $r_c(B_0,P,\theta,m_a) > R_{\text{NS}}$ in Eq.~(\ref{eq:rc}). The fact that $r_c \propto m_a^{-2/3}$ and $m_a \propto 1/F_a$ makes the previous condition easy to satisfy for large axion decay constants in most part of the $\theta$-parameter space.\footnote{There exists a set of polar angles, ${\Theta}$, which satisfy $r_c(\theta)>R_{\text{NS}}$ as shown Eq.(5.10)-(5.12) in ~\cite{Nurmi:2021xds}. As the axion mass increases, the angular regions associated with a null resonant axion-photon conversion slowly grow until
they eventually extend over the whole angular space, i.e. $\Theta = \{\}$.}

Eq.~(\ref{eq:S}) needs to be compared with the
minimum detectable flux of the chosen radio telescope, i.e.
 \begin{equation}
 \hspace{-1cm}S_{\text{min}}=\text{SNR}_{\text{min}} \frac{\text{SEFD}}{\eta_s \sqrt{2 \Delta B \Delta t_{\text{obs}}}}\hspace{0.5cm}\,,\label{radiometereq}
 \end{equation}
 \vspace{-0.5 cm}
 \begin{equation}
 \begin{adjustbox}{max width=218pt}
$
  \sim 220\, \mu\text{Jy}\left( \frac{\text{SNR}_{\text{min}}}{5} \right)\left( \frac{\text{SEFD}}{10\,\text{Jy}} \right) \left(  \frac{0.9}{\eta_s}\right) \left( \frac{1\,\text{kHz}}{\Delta B} \right)^{1/2} \left( \frac{1\,\text{yr}}{\Delta t_{\text{obs}}} \right)^{1/2}\,,
  $
\end{adjustbox}
 \end{equation}
 where $\text{SNR}_{\text{min}}$ is the minimum signal-to-noise ratio,  $\text{SEFD} \equiv 2 k_{\text{B}} T_{\text{sys}}/A_{\text{e}}$ is the so-called system equivalent flux density with $T_{\text{sys}}$ and $A_{\text{e}}$ the system temperature and the effective area,  respectively, $\Delta B$ is the telescope bandwidth, $\eta_s$ is the efficiency of the system, and $\Delta t_{\text{obs}}$ is the time of observation.
 
 The Square Kilometre Array (SKA) covers frequencies from 50 MHz to 350 MHz (SKA1-low) and 350 MHz to 14 GHz (SKA-mid). While the SKA1-low shows a characteristic SEFD, sensitivity, and spectral resolution of $2.8\,\text{Jy}$, $103\, \mu\text{Jy}-\text{hr}^{-1} (100\,\text{kHz}/\Delta B)^{1/2}$, and $1\,\text{kHz}$, respectively, the SKA-mid holds a characteristic SEFD, sensitivity, and spectral resolution of  $1.7\,\text{Jy}$, $63\, \mu\text{Jy}-\text{hr}^{-1} (100\,\text{kHz}/\Delta B)^{1/2}$, and $3.9\,\text{kHz}$, respectively (in both cases, the sentivity is calculated taking $\text{SNR}_{\text{min}}=\eta_s=1$)~\cite{SKAGuide}. 
\begin{figure}[t!]
\centering
\hspace*{-5mm}
\includegraphics[width=0.4\textwidth]{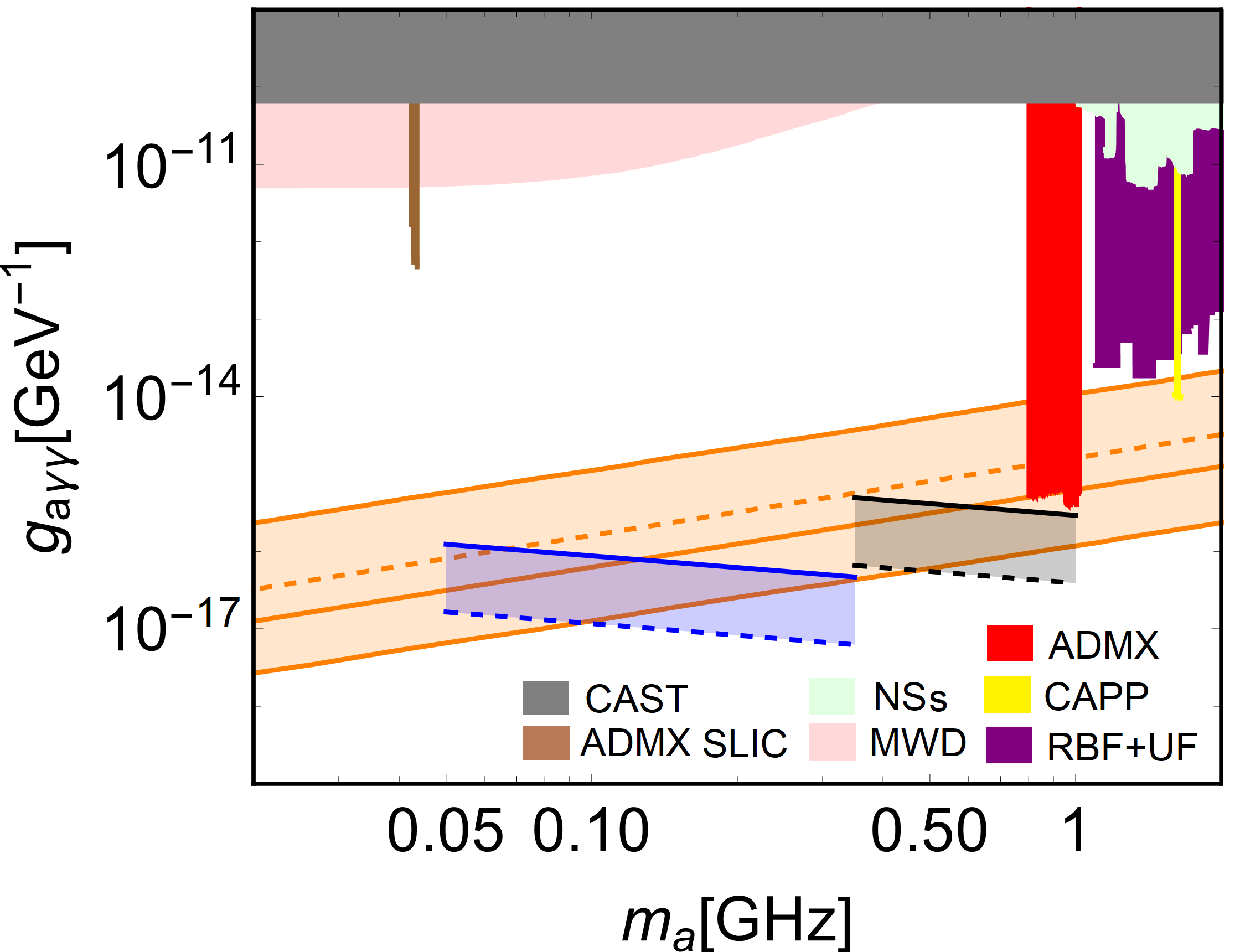}
\caption{Projected sensitivity for the axion-photon coupling constant associated with a NS-UMCH encounter. The blue (black) lines use SKA1-low array (SKA-mid array) with $\text{SNR}_{\text{min}}=5$, efficiency $\eta_s = 1$, and $\Delta B=\Delta \nu$. The rest of specifications are given in Table~\ref{tab:gayy}. For all cases, we take $M_{\text{UCMH}}=10^2 \,M_{\text{PBH}}$ and $M_{\text{PBH}}=M_{\odot}$. The orange band is the QCD axion parameter of space, where we have marked the KSVZ (dashed line) and DFSZ (solid line) models. Current constraints on the axion-photon coupling constant  are indicated with the remaining colored areas (see main text).}
\vspace*{-1.5mm}
\label{fig:gayy}
\end{figure}
\begin{center}
\begin{table}
\begin{tabular}[c]{|>{\columncolor[gray]{0.9}} c| c| c |}
\hline
 Array & SKA1-low & SKA-mid \\ 
 \hline
 P & 0.5 s & 0.5 s \\  
 \hline
 $B_0$ & $\mathcal{O}(10^{12})\,\text{G}$ & $\mathcal{O}(10^{13})\,\text{G}$\\
 \hline
 $\theta$ & $\pi/4$ & $\pi/4$\\
 \hline
  $\theta_m$ & $\pi/180$ & $\pi/180$\\
  \hline
 d & 300\,\text{pc}&\text{10\,kpc}\\
 \hline
 $\Delta t_{\text{obs}}$ & 100\,\text{hr}&100\,\text{hr}\\
 \hline
 $\Delta \nu$ & 30\,\text{kHz}& 100\,\text{kHz}\\
 \hline
 r (solid) & $10^{-5}\,R_{\text{UCMH}}$ &$10^{-5}\,R_{\text{UCMH}}$\\
 \hline
 r (dashed) &$10^{-6}\,R_{\text{UCMH}}$&$10^{-6}\,R_{\text{UCMH}}$\\
 \hline
\end{tabular}
\caption{\label{tab:gayy}Parameters used to calculate the projected sensitivity for the axion-photon coupling constant in Fig.~\ref{fig:gayy} (blue and black dashed and solid lines).}
\end{table}
\end{center}

 Galactic pulsar surveys and simulations show that NSs spin period and magnetic field at poles obey both log-normal distributions, with a mean and dispersion given by
 $\text{log}_{10}(P/ms) = 2.7$, 
$\sigma_{P} = -0.34$,~\cite{Lorimer2006qs} and
$\text{log}_{10}(B_0/G) = 12.65$, $\sigma_{B_{0}} = 0.55$,~\cite{FaucherGiguere2005dxp, Bates2013uma}, respectively. Based on that, we consider NS-UCMH encounters involving typical galactic NSs with spin periods and magnetic fields around mean values. For a small misalignment angle in Eq.~(\ref{eq:B}), such mean values lead to a broadened signal $\Delta \nu \sim (10-100)\,\text{kHz}$. We fix the
signal bandwidth received by SKA1-low and SKA-mid radio telescope to be $30\,\text{kHz}$ and $100\,\text{kHz}$, respectively. As we explain later, for a given axion mass and NS properties,  we choose the NS magnetic 
field at poles which is consistent with the predefined signal bandwidth. 

Fig.~\ref{fig:gayy} shows the projected sensitivity of the axion-photon coupling constant based on a NS-UCMH encounter by using the SKA1-low (blue lines) and SKA-mid (black lines) array. In all cases, we have taken
$M_{\text{UCMH}}=10^2\,M_{\text{PBH}}$ with $M_{\text{PBH}}=M_{\odot}$, $M_{\text{NS}}=M_{\odot}$, and $R_{\text{NS}}=10\,\text{km}$. The NS properties and other parameters are given in detail in Table 1. While the orange band indicates the parameter space for the QCD axion, the other colored regions correspond to the different constraints over the axion-photon coupling constant~\cite{AxionLimits} coming from CAST~\cite{CAST:2007jps, CAST:2017uph} (gray), ADMX~\cite{ADMX:2021nhd} (red),  ADMX SLIC~\cite{Crisosto:2019fcj} (brown), CAPP~\cite{Lee:2020cfj} (yellow), RBF+UF~\cite{PhysRevLett.59.839, PhysRevD.42.1297} (purple), Neutron Stars~\cite{Foster:2020pgt, Darling:2020plz, Darling:2020uyo} (green), and MWD~\cite{Dessert:2022yqq} (pink). 
The spectral density is estimated by
using Eq.~(\ref{eq:S}), where the axion density at the conversion radius is calculated from Eq.~(\ref{eq:rhorc}). For a given axion mass, misalignment angle, and spin period, we set the NS magnetic field at poles by requiring $\Delta \nu(B_0) = \Delta B$, assuming that the optimized telescope bandwidth matches the signal bandwidth in  Eq.~(\ref{eq:B}). We see that for $r\sim(10^{-6}-10^{-5})\,\text{pc}$ and an observation time $\Delta t_{\text{obs}}=100\,\text{hr}$, the SKA telescope  has the enough sensitivity to detect the QCD axion (orange band) from a source located at $\sim(0.1-10)\,\text{kpc}$. 

\textcolor{black}{
Here we have considered typical parameter values for neutron stars in the Galaxy. However, the effect of their values on the QCD axion detection can readily be extracted by using Eqs.\,(\ref{eq:S}) and (\ref{eq:B}). In our estimation, for fixed observation angle and $\Delta B=\Delta \nu$, the minimun axion-photon coupling to be detectable for a given telescope array reads as 
\begin{equation}
g_{a\gamma\gamma, \text{min}} \propto B_0^{-1/3} P^{-11/12} R_{\text{NS}}^{-1} M_{\text{NS}}^{-1/4}.
\end{equation}
Generally speaking, we see that the larger the magnetic field or the longer the NS spin period, the better is the sensitivity. From Fig.~\ref{fig:gayy}, for example, we see that even neutron stars holding low magnetic fields as $B_0 \sim 10^{10}\,\text{G}$ or short spin periods as $P \sim 0.1\,\text{s}$ should still lead to a detectable signal during a close encounter with an UCMH.}

From the perspective of axion searches, the analyzed range for axion masses complements the current parameter space associated with the axion-photon coupling constant. In addition, our proposed scenario can be also studied in the context of multi-messenger astronomy by considering the situation where the NS undergoes an inspiral motion to the central PBH leading to the associated gravitational wave emission. Such a situation was analyzed by~\cite{Edwards:2019tzf} in the context of intermediate black holes $(10^3-10^5\,)M_{\odot}$ having axion DM spikes around them. We leave this fascinating research avenue for a future work. 

\section{Conclusion}
\label{sec:conclusion}
In this work, we considered the axion DM scenario with the decay constant range $F_{a}\in[10^{12}{\rm GeV},10^{16}{\rm GeV}]$. Motivated by the story of axion embedded in either of supersymmetric models or string theory, we assumed a heavy scalar field $\sigma$ (saxion or moduli field) and studied the relic abundance of the axion DM during the EMD era in Sec.~\ref{sec:axionabundance} and \ref{sec:ADM}. The presence of $\sigma$ was shown to lead to the EMD era and its decay prior to BBN era causes the late time entropy production. Thanks to the later, even for $\theta_{\rm i}=\mathcal{O}(1)$, the axion DM relic abundance was shown to be consistent with that of the current DM, i.e. $\Omega_{\rm DM}h^{2}\simeq0.12$. 

Keeping in mind the presence of the EMD era as the advantageous environment for formation of PBH, we considered an enhanced power spectrum of the primordial curvature perturbation at $k\gtrsim10^{4}{\rm Mpc}^{-1}$. In Sec.~\ref{sec:PBH}, we showed how the assumed enhancement consistent with the constraint on $P_{\zeta}(k)$ could result in the formation of PBH with $f_{\rm PBH}=\mathcal{O}(10^{-3})$ for $M_{\rm PBH}\lesssim10^{-6}M_{\odot}$ (case I) and $M_{\rm PBH}=\mathcal{O}(1)M_{\odot}$. PBHs so formed are expected to develop to UCMH with $M_{\rm UCMH}\simeq10^{2}M_{\rm PBH}$ and $f_{\rm UCMH}\simeq10^{2}f_{\rm PBH}$ as was discussed in Sec.~\ref{sec:UCMH2}.

In Sec.~\ref{sec:exp} and \ref{sec:caseII}, we discussed how the UCMH can be of help in searching for the large decay constant axion DM. Firstly for the case I, UCMH can go through tidal disruption on high-speed encounters with stars. We computed the critical impact parameter $b_{c}$ numerically based on which we estimated the total probability for UCMH disruption. We found that for $M_{\rm PBH}\lesssim10^{-6}M_{\odot}$, all UCMHs go through disruption. The disruption gives rise to the tidal stream with $\rho_{\text{st}}$ larger than the current local DM density, which could potentially improve the experimental sensitivity of ABRACADABRA~\cite{Kahn:2016aff} for $g_{a\gamma\gamma}$ during the time of the encounter between the tidal stream and the earth.  Particularly for $M_{\rm PBH}$ as small as $\mathcal{O}(10^{-15})M_{\odot}$,  the Earth can encounter the tidal stream of axion DM with the enhanced $\rho_{\text{st}}\gtrsim 10\,\rho_{\text{local}}$ for $M_{\text{PBH}}\lesssim 10^{-14}M_{\odot}$ once every (1-10) years (see Fig.~\ref{fig:Gamma}).

For the case II, we attended to the possibility of transient radio signal generation when solar mass UCMHs encounter NSs. At the conversion radius $r_{c}$, the axion halo is expected to be converted into photons via the inverse Primakoff effect provided $m_{a}$ coincides with the plasma frequency. By considering (1) the Doppler broadening caused by temporal dependence from the co-rotation
of the plasma with the NS, (2) the Liouville's theorem for the axion density at the NS conversion radius, (3) typical NSs in the galaxy, and
(4) an observation time of $100 \,\text{hr}$,
we showed that the SKA1-low and SKA-mid  telescope have the
enough sensitivity to probe the parameter space $(m_{a},g_{a\gamma\gamma})$ as shown in Fig.~\ref{fig:gayy}.

In sum, we considered the non-standard cosmology with the EMD era as a way to save the axion DM scenario with $F_{a}\in[10^{12}{\rm GeV},10^{16}{\rm GeV}]$ from the dangerous overclosure of the universe. With the assumption for the enhanced $P_{\zeta}(k)$ on small scales, significant amount of PBHs can form during the EMD era, resulting in formation of UCMH during the late time matter domination era
via secondary infall accretion. We point out that tidal streams and transient radio signals generated on UCMH's encountering stars and NSs respectively can be possibly very helpful in searching for the large decay constant axion DM. Although not probed in this work, the induced gravitational wave due to the enhanced $P_{\zeta}(k)$ on small scales can be an interesting complementary experimental way to test our scenario~\cite{Inomata:2017uaw}. 

\vspace{-1 cm}
\begin{acknowledgments}
This work was supported by the Academy of Finland grant 318319. G.C. thanks Emilian Dudas and Keisuke Harigaya for useful discussion about the string axion, and Juan Garcia-Bellido for email conversation about constraints on the primordial power spectrum, and Fabrizio Rompineve for discussion for PBH. E.D.S. thanks Tsutomu T. Yanagida (Shanghai Jiao Tong University) and his comment about the effect of the central PBH mass in the UCMH resistance against disruption during high speed encounters with stars. E.D.S. thanks Björn Garbrecht (Technische Universit$\ddot{a}$t M$\ddot{u}$nchen), Jamie McDonald (Catholic University of Louvain), and Sankarshana Srinivasan
(University of Manchester) for useful discussion about the axion broadening signal in the magnetosphere of neutron stars. E.D.S. thanks Thomas Edward (Stockholm University) and Sami Nurmi (University of Jyv$\ddot{a}$skyl$\ddot{a}$ and University of Helsinki) for discussion about the axion density at the conversion radius in the neutron star magnetosphere. 
\end{acknowledgments}
\vspace{2 cm}
\appendix
\section{Hubble Expansion Rate at $t_{*}$}
\label{appA}
Define $t_{*}$ ($R_{*}$) to be the time (the scale factor) at which the EMD era gets started.  Assume that the heavy particle $\sigma$ starts oscillation at the time of rh1. Then from the equality $\rho_{\sigma}(R_{*})\simeq\rho_{\rm rad}(R_{*})$, we obtain
\beq
\frac{1}{2}m_{\sigma}^{2}\sigma_{0}^{2}\left(\frac{R_{\rm rh1}}{R_{*}}\right)^{3}\simeq g_{*}(T_{\rm rh1})\frac{\pi^{2}}{30}T_{\rm rh1}^{4}\left(\frac{R_{\rm rh1}}{R_{*}}\right)^{4}\,,
\label{eq:rhosigma}
\eeq
where $\sigma_{0}$ is the initial field displacement from the origin in the field space, $m_{\sigma}$ is the mass of $\sigma$ and $T_{\rm rh1}$ is the reheating temperature defined via the inflaton perturbative decay rate $T_{\rm rh1}\simeq\sqrt{\Gamma_{\rm inflaton}M_{P}}$. 

The ratio $R_{\rm rh1}/R_{*}$ being inferred from Eq.~(\ref{eq:rhosigma}), we can $\rho_{\sigma}(R_{*})$
\beq
\rho_{\sigma}(R_{*})=\left(\frac{15^{3}}{2\pi^{6}g_{*}(T_{\rm rh1})^{3}}\right)\frac{m_{\sigma}^{8}\sigma_{0}^{8}}{T_{\rm rh1}^{12}}\,.
\label{eq:rhosigmainitial}
\eeq
which in turn gives
\beq
H(R_{*})=\sqrt{\frac{15^{3}}{6\pi^{6}g_{*}(T_{\rm rh1})^{3}}}\frac{m_{\sigma}^{4}\sigma_{0}^{4}}{M_{P}T_{\rm rh1}^{6}}\,.
\label{eq:Ha*}
\eeq
Since we assumed that the moduli starts oscillation at rh1, we have
\beqs
m_{\sigma}^{2}&\simeq&H(a_{\rm rh1})^{2}=\frac{g_{*}(T_{\rm rh1})\frac{\pi^{2}}{30}T_{\rm rh1}^{4}}{3M_{P}^{2}}\cr\cr&\rightarrow&\quad T_{\rm rh1}^{6}=m_{\sigma}^{3}\left(\frac{g_{*}(T_{\rm rh1})\frac{\pi^{2}}{30}}{3M_{P}^{2}}\right)^{-3/2}\,.
\label{eq:Trh1}
\eeqs
Finally by plugging Eq.~(\ref{eq:Trh1}) in Eq.~(\ref{eq:Ha*}), we get
\beq
H(R_{*})\equiv H_{*}=0.765m_{\sigma}\left(\frac{\sigma_{0}}{M_{P}}\right)^{4}\,.
\label{eq:Hinitial}
\eeq

\section{PBH Mass Range ($M_{\rm max}$ and $M_{\rm min}$)}
\label{appB}
When the overdensity associated with the wavenumber $k=R_{k}H(R_{k})$ goes through the gravitational collapse to form a PBH, the PBH's mass is given by
\beq
M=\gamma M_{H}=\gamma\frac{4\pi M_{P}^{2}}{H}\,.
\label{eq:PBHmass}
\eeq
where $M_{H}=(4\pi/3)H^{-3}\rho_{\rm tot}$ is the horizon mass and $\gamma=1$ if $k$ re-entered the horizon at a matter dominated era. By replacing $H$ with $k/R_{k}$ and writing $R_{k}$ in terms of $R_{\rm rh2}$ and $T_{\rm rh2}$ based on the entropy conservation, one can find that PBH mass is mapped to $k$ via
\beqs
M&=&\gamma\frac{2\pi^{3}}{45}\left(\frac{T_{0}}{k}\right)^{3}\frac{g_{s*}(T_{0})}{g_{s*}(T_{\rm rh2})}g_{*}(T_{\rm rh2})T_{\rm rh2}\cr\cr
&\simeq&2.42M_{\odot}\times\left(\frac{k}{10^{6}{\rm Mpc}^{-1}}\right)^{-3}\left(\frac{T_{\rm rh2}}{10{\rm MeV}}\right)\,.\nonumber \\
\label{eq:MPBH}
\eeqs
On observing Eq.~(\ref{eq:MPBH}), it is realized that the minimum PBH mass $M_{\rm min}$ is associated with $k_{\rm min}$ re-entering the horizon at $R_{*}$ while $M_{\rm max}$ is so with $k_{\rm max}$ of which corresponding fluctuation grows to 1 at $R_{\rm rh2}$.

When the Fourier modes of primordial fluctuations lying in $k_{\rm rh2}<k<k_{\rm min}$ re-entered the horizon, the universe was in the EMD era. Both of $k_{\rm min}$ and $k_{\rm max}$ can be expressed in terms of $k_{\rm rh2}$. Aiming to calculate $k_{\rm min}$ and $k_{\rm max}$, we first compute $k_{\rm rh2}=R_{\rm rh2}H_{\rm rh2}$. From the entropy conservation, we have
\beqs
R_{\rm rh2}&=&R_{\rm eq}\frac{g_{s*}(R_{\rm eq})^{\frac{1}{3}}T_{\rm eq}}{g_{s*}(R_{\rm rh2})^{\frac{1}{3}}T_{\rm rh2}}\cr\cr
&\simeq&1.58\times10^{-11}\left(\frac{T_{\rm rh2}}{10{\rm MeV}}\right)^{-1}\,,
\label{eq:arh2}
\eeqs
where $g_{s*}(R_{\rm eq})=3.94$ and $g_{s*}(R_{\rm rh2})=10.75$ were used. Along with $H_{\rm rh2}=\Gamma_{\sigma}$ from Eq.~(\ref{eq:saxiondecayrate}), Eq.~(\ref{eq:arh2}) gives
\beqs
k_{\rm rh2}&\simeq&4.3\times10^{5}{\rm Mpc}^{-1}\times\left(\frac{T_{\rm rh2}}{10{\rm MeV}}\right)^{-1}\left(\frac{m_{\sigma}}{100{\rm TeV}}\right)^{3}\,.\nonumber \\
\label{eq:krh2}
\eeqs

On the other hand, from $k_{\rm min}/k_{\rm rh2}=(R_{*}H_{*})/(R_{\rm rh2}H_{\rm rh2})=(R_{\rm rh2}/R_{*})^{1/2}$, we can write $k_{\rm min}$ as
\beqs
k_{\rm min}&=&k_{\rm rh2}\times\left(\frac{R_{\rm rh2}}{R_{*}}\right)^{1/2}\cr\cr
&=&k_{\rm rh2}\times\left(\frac{H_{*}}{H_{\rm rh2}}\right)^{1/3}\cr\cr
&=&k_{\rm rh2}\times(0.765m_{\sigma})^{1/3}\left(\frac{\sigma_{0}}{M_{P}}\right)^{4/3}\Gamma_{\sigma}^{-1/3}\,,\nonumber \\
\label{eq:kmin}
\eeqs
where we used $R_{\rm rh2}/R_{*}=(t_{\rm rh2}/t_{*})^{2/3}=(H_{*}/H_{\rm rh2})^{2/3}$ for the second equality, and $
H_{\rm rh2}\simeq\Gamma_{\sigma}$ and $H_{*}$ in Eq.~(\ref{eq:Hinitial}) for the last equality. Therefore, plugging Eq.~(\ref{eq:krh2}) in Eq.~(\ref{eq:kmin}) yields
\beqs
k_{\rm min}&\simeq&3.274\times10^{14}{\rm Mpc}^{-1}\cr\cr&\times&\left(\frac{T_{\rm rh2}}{10{\rm MeV}}\right)^{-1}\left(\frac{\sigma_{0}}{M_{P}}\right)^{\frac{4}{3}}\left(\frac{m_{\sigma}}{100{\rm TeV}}\right)^{\frac{7}{3}}\,.\nonumber \\
\label{eq:kmin2}
\eeqs

As for $k_{\rm max}$, given that the primordial fluctuation associated with the wavenumber $k_{\rm max}$ grows to 1 after the horizon re-entry at $R_{\rm rh2}$, we have $\sigma\times(R_{\rm rh2}/R_{\rm hc})=1$. Here $k_{\rm max}$ is assumed to re-enter the horizon at $R_{\rm hc}$. Because of $R_{\rm rh2}/R_{\rm hc}=(H_{\rm hc}/H_{\rm rh2})^{2/3}$, this in turn gives
\beq
H_{\rm hc}\simeq\sigma^{-3/2}H_{\rm rh2}\,.
\label{eq:Hhor}
\eeq
Hence, now that $M\propto k^{-3}$ and $M\propto H^{-1}$, we obtain
\beq
k_{\rm max}=k_{\rm rh2}\sigma_{\rm max}^{-1/2}\,.
\label{eq:kmax}
\eeq
Note that $\sigma_{\rm max}$ can be obtained from $\sigma_{\rm max}^{2}\simeq(2/5)^{2}P_{\zeta}(k_{\rm max})$ and it reads~\cite{Bhattacharya:2021wnk}
\vspace{1 cm}
\beqs
\log\sigma_{\rm max}&=&2\log\left(\frac{k_{\rm rh2}}{k_{p}}\right)-8\sigma_{p}^{2}\cr\cr&+&2\sqrt{16\sigma_{p}^{4}-8\sigma_{p}^{2}\log\left(\frac{k_{\rm rh2}}{k_{p}}\right)+2\sigma_{p}^{2}\log\left(\frac{4A_{p}}{25}\right)}\,.\nonumber \\
\label{eq:sigmamax}
\eeqs

\section{More accurate UCMH profile}
\label{UCMHaccurate}

Under the approximation that the DM velocity dispersion at UCMH formation $\sigma_{\text{DM}}$ is not affected by the UCMHs themselves, the mean tangential velocity of the infalling DM particle as a function of the radius ($r$) and the redshift ($z$) is
\begin{align}
&v_{\text{rot}}(r,z) = \frac{\sigma_{\text{DM}}(z)R_{\text{UCMH}}(z)}{r}\approx 1.4\times10^{-4}\,\text{km/s}\, \times\nonumber\,\\
&\left( \frac{1+z}{1000} \right)^{-1/2} \left( \frac{M_{\text{UCMH}}(z)}{M_{\odot}} \right)^{0.28} \frac{R_{\text{UCMH}}(z)}{r}\,,     
\end{align}
where we have used Eq. (19) in~\cite{Ricotti:2007au} and $R_{\rm UCMH}(z)$ in Eq.~(\ref{rhalo}).  

The radial infall regime breaks down when the tangential velocity of a DM particle which is falling at radius $r$ from the central PBH overpasses the local Keplerian orbital velocity
\begin{equation}
v_{\text{kep}}(r) =  \frac{\sqrt{G_N M_{\text{UCMH}}(z)}}{r^{1/8}R_{\text{UCMH}}^{3/8}(z)}\,. 
\end{equation}
 Equating $v_{\text{rot}}(r_c) = v_{\text{kep}}(r_c)$ yields 
\begin{equation}
r_{\text{c}} \approx 6\times 10^{-7}\text{pc} \left( \frac{1000}{z_{\sigma}+1} \right)^{2.415} \left( \frac{M_{\text{PBH}}}{M_{\odot}} \right)^{0.272}\,,   
\end{equation}
which is in close agreement with~\cite{2012PhRvD..86d3519L}. Here
$r_c$ is the radius of the non-radial infall core and $z_{\sigma}$ is the redshift at which the UCMH collapses~\cite{Ricotti:2009bs}.  Typical conservative estimate obeys the parametrization~\cite{2012PhRvD..86d3519L}
\begin{equation}
\rho_{\text{UCMH}} = \rho_{\text{UCMH}}(r_c) \left( 1 + \frac{r}{r_c} \right)^{-9/4}\,.     
\end{equation}

\section{Ergodic Distribution Function for UCMHs}
\label{EGDF}
Here we determine the ergodic distribution function (DF)~\footnote{For a reivew of this topic, we refer the readers to Chapter 4 in Ref.~\cite{2008gady.book.....B}.} associated with the UCMH spherically symmetric radial profile. Knowing the DF will allow us to determine the axion velocity distribution with respect to the UCMH radius. 
The UCMH radial profile is simply written as $\rho_{\text{UCMH}}(r)=\rho_R(R_{\text{UCMH}}/r)^{\gamma}$, where 
$\rho_R \equiv \rho_{\text{UCMH}}(R_{\text{UCMH}})$ and  $\gamma=9/4$, Eq.~(\ref{infall}). The enclosed UCMH mass at radius $r$ reads
\begin{align}
M_{\text{UCMH}}(r) &\approx 4\pi \int_0^{r} r'^2 \rho_{\text{UCMH}}(r')dr' \,,
\\
& \approx\frac{16\pi \rho_{R}R_{\text{UCMH}}^3}{3}\left( \frac{r}{R_{\text{UCMH}}}\right)^{3/4}\,,
\end{align}
\textcolor{black}{where the approximation symbol comes from the fact that we have extended the lower limit of the integral to zero and we have also neglected the formation of a core at very small radius~\cite{Bringmann:2011ut}}.
We define the UCMH relative potential as $\Psi(r)=-\Phi(r)+\Phi_0$ 
where $\Psi \rightarrow \Phi_0=0$ as $r\rightarrow \infty$. 
Under the approximation $M_{\text{UCMH}}(r) + M_{\text{PBH}} \approx M_{\text{UCMH}}(r)$ for sufficiently large $r$, we have
 \begin{equation}
\rho_{\text{UCMH}}(\Psi)=\left(\frac{3\Psi}{16\pi G_N \rho_R^{8/9}R_{\text{UCMH}}^2} \right)^9\,,\label{C4}
 \end{equation}
 where we have used $\Psi(r)=G_N M_{\text{UCMH}}(r)/r$\,.
Noting that $(d\rho_{\text{UCMH}}/d\Psi)|_{\Psi=0}=0$ and defining the relative energy $\xi = \Psi - \frac{1}{2}v_a^2$, with $v_a$ as the axion velocity,
the (simplified) Eddington's inversion formula for the DF reads as
\begin{equation}
f(\xi) = \frac{1}{\sqrt{8}\pi^2} \left[  \int_0^{\xi} \frac{d^2\rho_{\text{UCMH}}}{d\Psi^2} \frac{d\Psi}{\sqrt{\xi-\Psi}} \right]\,,\label{C5}
\end{equation}
Plugging Eq.~(\ref{C4}) into Eq.~(\ref{C5}), we find
\begin{align}
    f(\xi) &=    \frac{18\sqrt{2} \rho_{\text{UCMH}}(\Psi)}{\Psi^9\pi^2}\int_0^{\xi}\frac{\Psi^{7}}{\sqrt{\xi-\Psi}}d\Psi\,,\\
&= \frac{A}{G_N^{9}\rho_{R}^8R_{\text{UCMH}}^{18}}\xi^{15/2}\,,\label{fepsilon}
\end{align}
where $A=(2998927360\sqrt{2}\pi^{11}/19683)^{-1}$. Then, the axion speed distribution at a radius $r$
is written as
\begin{align}
    f(v_a|r) &= 4\pi v_a^2 \frac{f(\Psi(r)-\frac{1}{2}v_a^2)}{\rho_{\text{UCMH}}(r)} \,,\\
    &= \frac{32768 \sqrt{2}}{715\pi} \frac{v_a^2(\Psi(r)-\frac{1}{2}v_a^2)^{15/2}}{\Psi^9(r)}\,,
\end{align}
with the normalization
\begin{equation}
    \int_{0}^{\sqrt{2\Psi(r)}} f(v_a|r)dv_a = 1\,.\label{eq:norm}
\end{equation}
The axion velocity at a given radius within the UCMH can be estimated by using the dispersion velocity $\sigma(r)$ from the distribution function $f(v_a|r)$ as
\begin{equation}
\sigma^2(r) = \int_0^{\sqrt{2\Psi(r)}}v_a^2f(v_a|r)dv_a = 0.3\Psi(r)\,.\label{eq:sigmaa}     
\end{equation}
\\
By using Eq.~(\ref{infall}), it is easy to show that the UMCH radius which encloses k-times the mass of the central PBH is estimated to be $r = R_{\text{UCMH}}(k M_{\text{PBH}}/M_{\text{UCMH}})^{4/3}$. For $M_{\text{UCMH}} = 10^{2}\,M_{\text{PBH}}$, we have $r = 2\times 10^{-3}R_{\text{UCMH}}$ (see Sec. II in ~\cite{Hertzberg:2020hsz}). As a result, for small radius $r$ such that $r \lesssim 2\times10^{-3}\,R_{\text{UCMH}}$, we can make the approximation $M_{\text{UCMH}}(r) + M_{\text{PBH}} \approx M_{\text{PBH}}$. 
We thus write the relative potential as $\Psi(r) \approx \Psi_{\text{PBH}}(r) = G_N M_{\text{PBH}}/r$ so that the density is re-expressed in terms of that potential as
\begin{equation}
\rho_{\text{UCMH}}(\Psi_{\text{PBH}}) = \rho_{R}\left( \frac{R_{\text{UCMH}}}{G_N M_{\text{PBH}}}\right)^{9/4}\Psi_{\text{PBH}}^{9/4}\,.    
\end{equation}
Following the same procedure as before, the axion DM speed distribution at a radius $r$ reads as
\begin{align}
&f(\xi) = \frac{45\Gamma[5/4]\rho_R}{16(2\pi)^{3/2}\Gamma[7/4]} \left( \frac{R_{\text{UCMH}}}{G_N M_{\text{PBH}}} \right)^{9/4} \xi^{3/4}\,,\\ 
&f(v_a|r)=\frac{45\Gamma[5/4]}{4\sqrt{8\pi}\Gamma[7/4]}\frac{v_a^2(\Psi_{\text{PBH}}(r)-\frac{1}{2}v_a^2)^{3/4}}{\Psi_{\text{PBH}}(r)^{9/4}}\,.\label{eq:fMPBH}
\end{align}
where the same normalization applied in Eq.~(\ref{eq:norm}) holds.

\bibliography{main}
\end{document}